\documentclass{aa}  

\usepackage{graphicx}
\usepackage{txfonts}
\usepackage{orcidlink}
\usepackage{booktabs}
\usepackage{upgreek}
\usepackage{threeparttable}
\usepackage[T1]{fontenc}
\usepackage{comment}
\DeclareRobustCommand{\VAN}[3]{#2}
\let\VANthebibliography\thebibliography
\def\thebibliography{\DeclareRobustCommand{\VAN}[3]{##3}\VANthebibliography}
\usepackage{graphicx}	% Including figure files
\usepackage{amsmath}	% Advanced maths commands
\usepackage{tablefootnote}
\usepackage{url}
\bibpunct{(}{)}{;}{a}{}{,} % to follow the A&A style
\usepackage{txfonts}
\usepackage{hyperref}
\hypersetup{colorlinks=true,linkcolor=blue,citecolor=blue,filecolor=blue}

\begin{document}

\title{MEDUSA}
\subtitle{I. Tracing magnetic field structures in tidal arms of the dwarf–dwarf merger NGC\,1487}
\titlerunning{MEDUSA I: Tracing magnetic field in NGC\,1487}

   \author{Sam Taziaux \inst{\ref{rub},\ref{rapp},\ref{csiro}}\orcidlink{0009-0001-6908-2433}
    \and
    Aritra Basu \inst{\ref{TLS}, \ref{mpi_bonn}}\orcidlink{0000-0003-2030-3394}
     \and 
    Samata Das \inst{\ref{tp4}, \ref{rapp}}\orcidlink{0009-0005-9140-7811}
    \and
    Dominik J. Bomans \inst{\ref{rub}, \ref{rapp}}\orcidlink{0000-0001-5126-5365}
    \and Timothy J. Galvin \inst{\ref{csiro}}\orcidlink{0000-0002-2801-766X}
     \and 
    Alec J. M. Thomson \inst{\ref{ska}}\orcidlink{0000-0001-9472-041X}
    \and 
    George H. Heald \inst{\ref{ska}}\orcidlink{0000-0002-2155-6054}
    \and
    Peter Kamphuis \inst{\ref{rub}}\orcidlink{0000-0002-5425-6074}
    \and
    Francesca Loi \inst{\ref{inaf}}\orcidlink{0000-0002-8627-6627}
    \and
    Michael Stein \inst{\ref{rub}}\orcidlink{0000-0001-8428-7085}
    \and
    Krysztof T. Chyży \inst{\ref{pol}}\orcidlink{0000-0002-6280-2872}
    \and
    Christopher J. Riseley\inst{ \ref{rub}, \ref{rapp}}\orcidlink{0000-0002-3369-1085}
    \and
    Ralf-Jürgen Dettmar \inst{\ref{rub}, \ref{rapp}}\orcidlink{0000-0001-8206-5956}
    \and
     Julia Becker Tjus\inst{\ref{tp4}, \ref{rapp},\ref{sweden}}\orcidlink{0000-0002-1748-7367}
          }

   \institute{Ruhr University Bochum, Faculty of Physics and Astronomy, Astronomical Institute (AIRUB), Universitätsstraße 150, 44801 Bochum, Germany
              \email{sam.taziaux@rub.de} \label{rub}
        \and 
             Ruhr Astroparticle and Plasma Physics Center (RAPP Center), 44780 Bochum, Germany \label{rapp}
     \and 
            CSIRO Space and Astronomy, PO Box 1130, Bentley WA 6102, Australia \label{csiro}
        \and 
             Th\"uringer Landessternwarte, Sternwarte 5, 07778 Tautenburg, Germany \label{TLS}
        \and
            Max-Planck-Institut für Radioastronomie, Auf dem H\"ugel 69, 53121 Bonn, Germany \label{mpi_bonn}
         \and
            Ruhr University Bochum, Faculty for Physics \& Astronomy, Theoretical Physics IV: Plasma Astroparticle Physics, 44780 Bochum, Germany \label{tp4}
        \and 
            SKA Observatory, SKA-Low Science Operations Centre, 26 Dick Perry Avenue,Kensington WA 6151, Australia \label{ska}
        \and 
            INAF – Osservatorio Astronomico di Cagliari, Via della Scienza 5, Selargius, Italy \label{inaf}
         \and 
            Astronomical Observatory of the Jagiellonian University, ul. Orla 171, 30-244 Kraków, Poland \label{pol}
        \and
            Department of Space, Earth and Environment, Chalmers University of Technology, Gothenburg, Sweden \label{sweden}
            }

% These dates will be filled out by the publisher
\date{Accepted XXX. Received YYY; in original form ZZZ}

% Enter the current year, for the copyright statements etc.
%\pubyear{2023}

% Don't change these lines
%\label{firstpage}

\abstract
{Radio continuum observations of dwarf--dwarf galaxy mergers, such as NGC 1487, provide crucial insights into magnetic field amplification and cosmic ray (CR) propagation during galactic assembly. Dwarf galaxies are important laboratories for studying cosmic magnetism because they can maintain strong magnetic fields via the action of turbulent dynamo despite their low mass and weak gravitational potential.}
{The Magnetic-field Evolution in Dwarf galaxies from Ultra-deep SKA Analysis (MEDUSA) survey is the first SKA-pathfinder programme designed to obtain deep continuum, polarisation, and H{\sc i} data for dwarf galaxies, enabling a comprehensive study of their radio spectra, magnetic fields, and gas kinematics across a representative population. By analysing the radio continuum spectra and polarisation of the dwarf-dwarf galaxy merger NGC\,1487 from the MEDUSA sample, we aim to determine its magnetic field strength and to characterise the large-scale and turbulent components of its magnetic field.}
{We utilise highly sensitive multi-band radio continuum data from MeerKAT L-band (1.28\,GHz) and Australia Telescope Compact Array (ATCA) L/S (2.1\,GHz), C (5.5\,GHz), and X-bands (9\,GHz). We analysed the magnetic field configuration using polarisation and rotation measure (RM) synthesis.}
{The integrated spectral energy distribution has a non-thermal spectral index of $\alpha_{\rm nth} = -0.678\pm0.085$, indicating a significant synchrotron contribution, consistent with a CR electron injection index of $\gamma = 2.36$ ($N(E) \propto E^{-\gamma}$) typical of supernova remnants. Synchrotron and inverse Compton losses cause a spectral break at $\nu_{\rm b} = 6.2\pm1.3$\,GHz. In star-forming regions, the magnetic field exhibits strong small-scale fluctuations in RM, suggesting the action of a small-scale dynamo.
Conversely, the field becomes more ordered, aligning with the tidal arms toward the galaxy's outskirts, showing a large-scale magnetic field over $\approx3$\,kpc. CR cooling timescale of approximately 11\,Myrs at 1.28\,GHz is similar to the escape timescale.}
{Observations of the dwarf–dwarf merger NGC\,1487 show that even low-mass galaxy mergers, likely the building blocks of larger galaxies in the early Universe, can rapidly amplify and produce coherent large-scale magnetic field structures, highlighting their key contribution in the early magnetisation of galaxies.}


   \maketitle
%


%%%%%%%%%%%%%%%%%%%%%%%%%%%%%%%%%%%%%%%%%%%%%%%%%%

%%%%%%%%%%%%%%%%% BODY OF PAPER %%%%%%%%%%%%%%%%%%

\section{Introduction}
Galaxy mergers are a fundamental process in the Universe, shaping the evolution of galaxies by driving mass assembly, altering morphology, and influencing star formation rates \citep{micic_two_2022, papaderos_dwarf_2012, Henkel_2022}. Among them, mergers between dwarf galaxies are particularly significant, as dwarf galaxies are the most common type \citep[e.g.][]{Henkel_2022} and serve as essential building blocks for larger systems. Their low mass and weak gravitational potential make them highly susceptible to  external influences such as tidal interactions, as well as to the effects of stellar feedback \citep[][]{Buzzo_2021, Kronberg_2016, Mengel_2008}. Compared to more massive galaxies, these dwarf galaxies rotate more slowly, reducing the efficiency of the $\Omega$-dynamo and hence the amplification of large-scale magnetic fields \citep[e.g.][]{Beck_2013}. At the same time, their shallow gravitational potential allows stellar feedback to launch comparatively stronger outflows \citep{Buzzo_2021}.
Even though their slow rotation and shallow gravitational wells would typically limit large-scale dynamo action, dwarf galaxies can still sustain strong magnetic fields powered by intense star formation and supernova-driven turbulence \citep[e.g.][]{chyzy_magnetized_2016, kepley_role_2010, Basu_2017, Kepley_2011}. This makes them key systems for investigating the fast magnetic field amplification processes believed to shape magnetic fields in the early Universe \citep{Schleicher_2013, Kronberg_2016, Rodenbeck_2016}.
Simulations show that magnetic field strength in galaxy mergers exhibits two peaks: the first during initial core approach and the second just before coalescence, driven by tidal forces and angular momentum transport \citep{Rodenbeck_2016}. In central regions, fields are amplified by angular momentum transport, while in outer regions, geometric projection effects enhance observed strength. Additionally, shear forces and Kelvin-Helmholtz instabilities in outflows further amplify magnetic fields \citep{Kronberg_2016}.

Radio continuum observations provide crucial key information into galactic magnetic fields, including their strength \citep{beck_2019} and orientation via polarisation \citep[][]{Brentjens_2005, Pakmor_2018, Reissl_2023}. Additionally, they serve as a powerful tool for studying cosmic ray (CR) transport and energy loss mechanisms (synchrotron, inverse-Compton, and bremsstrahlung) using synchrotron emission as tracers \citep[e.g.][]{Lacki_2010, Werhahn_2021, Pfrommer_2022, heesen_nearby_2022}. 
CRs produced in supernova remnants travel into the halo through advection, diffusion, and streaming \citep[e.g.][]{Strong_2007, Stein_2019, Thomas_2020}, exciting Alfvén and whistler waves that scatter them and regulate their drift speed \citep{Kulsrud_1969, Shalaby_2021, Shalaby_2023, Lemmerz_2024}. This coupling allows CRs to transfer momentum to the gas and generate pressure support that can drive large-scale galactic winds, a key feature of CR hydrodynamic models \citep{Zweibel_2013, Pfrommer_2017, Thomas_2019}. Magneto-hydrodynamic simulations have shown that such CR-driven winds can significantly influence mass loading, halo gas distribution, and star formation \citep[e.g.][]{Breitschwerdt_1991, Uhlig_2012, Pakmor_2016, Girichidis_2016, Recchia_2017, Dashyan_2020, Thomas_2023}. This study extends earlier radio continuum \citep[e.g.][]{chyzy_regular_2000, kepley_role_2010, chyzy_magnetized_2016,Kepley_2011, Basu_2017, taziaux_2025}, by using low-frequency observations to trace CR electrons (CREs) into galaxy halos and characterise their energy losses. CR protons dominate in dense gas through hadronic interactions \citep[][]{Pfrommer_2004, Pfrommer_gamma_2017, Werhahn_2023}, while synchrotron-emitting CREs probe lower-density outflows \citep{Ruszkowski_2023}. CRs can also advect magnetic fields into galaxy halos, an effect expected to be especially strong in dwarf galaxies with shallow gravitational potentials. However, detecting CR-driven winds in dwarfs is observationally difficult because their synchrotron halos are faint, low in surface brightness, and easily diluted by thermal emission \citep[e.g.][]{heesen_exploring_2018, taziaux_2025}. Low-frequency observations, with their enhanced sensitivity to older CRE populations, are therefore particularly well suited to testing CR feedback and its influence on magnetic fields.

In this paper, we present a full Stokes continuum study of the dwarf-dwarf merger NGC\,1487 to explore the outflow kinematics, magnetic field structure, the dependence of energy loss mechanisms and the CR transport in dwarf galaxy mergers. These datasets represent the most sensitive set of radio continuum observations of this low-mass merger NGC\,1487 to date and the first observations of their radio continuum and polarised emission at 1.28, 2.1, 5.5 and 9\,GHz. 
We describe our data set and reduction in Sect.~\ref{dataset}. Spectral properties will be analysed in Sect.~\ref{radiocont}. In Sect.~\ref{mfconfig}, we describe the magnetic field configuration based
on polarisation measurement and Faraday rotation. Further, in Sect.~\ref{crtransport}, the scope of modelling the CR transport in the scenario of NGC\,1487 is explored. We will discuss the role of the magnetic field and the CR transport in Sect.~\ref{discussion} and the work will be summarised in Sec.~\ref{conclusion}. Throughout the paper, we use the Gaussian cgs unit system.

\section{Observation and data reduction}
\label{dataset}

\subsection{Selection of target}
   \begin{figure*}
    \centering
    \includegraphics[width=\linewidth]{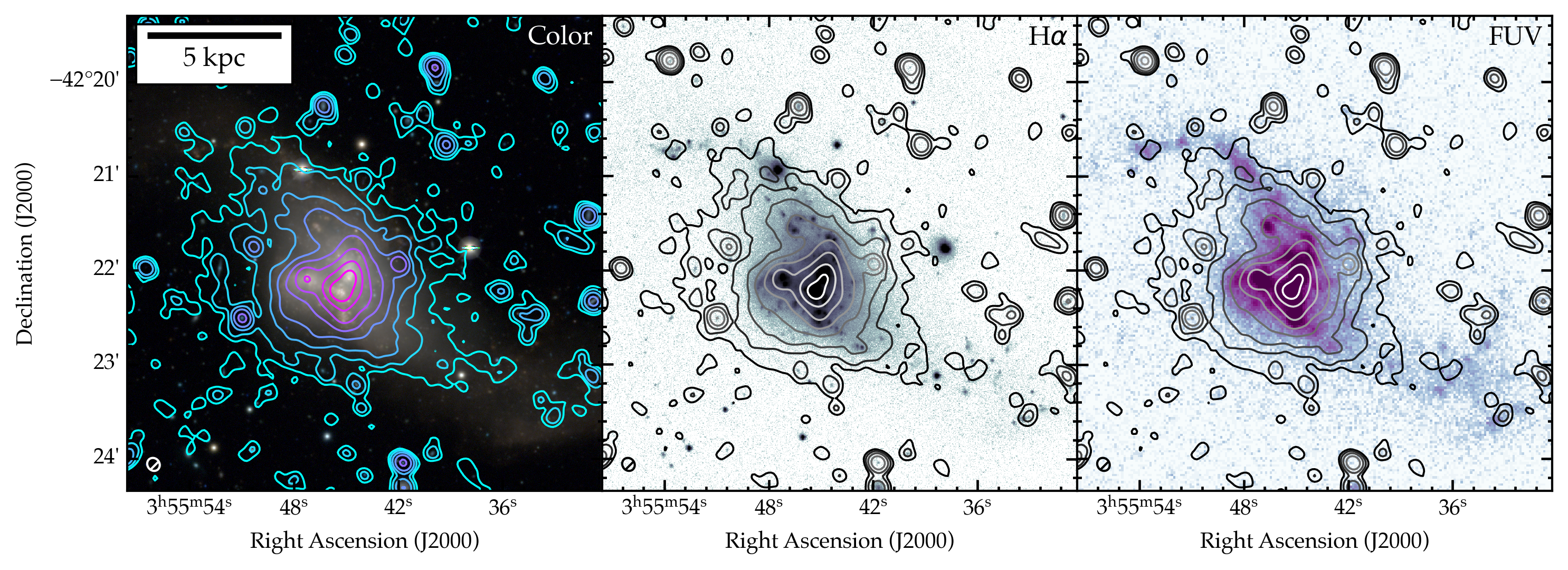}
    \caption{Colour-composite images from the DESI Legacy Imaging Surveys (\textit{left panel}), HIPASS H$\alpha$ image (\textit{middle panel}) and GALEX FUV image (\textit{right panel}) of NGC\,1487 with overlaid MeerKAT radio emission contours starting at $3\,\sigma_I$ and increasing by a factor of 2 at a central frequency of 1.28\,GHz ($\sigma_I=4\,\upmu$Jy/beam). The beam of $8.56\arcsec \times 7.65\arcsec$ is shown in the left corner. The scale is calculated using the distance given in Table~\ref{basics}.}
    \label{optical}
\end{figure*}

The Magnetic-field Evolution in Dwarf galaxies from Ultra-deep SKA Analysis (MEDUSA) survey is a radio polarimetry program (PI: S. Taziaux) using the MeerKAT telescope to study magnetic fields, neutral hydrogen, and CR transport in low-mass galaxies. The survey targets a representative sample of 13 nearby dwarf galaxies ($D \lesssim 13$\,Mpc) covering a broad range of morphologies, gas fractions, star-formation rates, and evolutionary stages. Systems include isolated dwarfs, and low-mass mergers, selected to enable resolved studies of synchrotron emission, depolarisation, and H{\sc i} kinematics. The full sample selection and survey strategy are described in Taziaux in preparation.

Within this study, NGC\,1487 was selected for an in-depth case study because it is a gas-rich, pre-merger dwarf galaxy \citep{Buzzo_2021}. In Fig~\ref{optical}, we show the optical, H$\alpha$ and FUV images overlaid with the MeerKAT L-band observations, revealing overlapping discs, tidal tails, and ongoing starbursts, characteristic of an early-stage merger \citep{Buzzo_2021, Bergvall1995}. Its tidal arms host active star formation, with star cluster mass distributions following a power law and star-formation rate densities of $(0.345 \pm 0.024) \times 10^{-3}\,{\rm M_\odot \,yr^{-1}\,kpc^{-2}}$ (east arm) and $(0.360 \pm 0.024) \times 10^{-3}\,{\rm M_\odot \,yr^{-1}\,kpc^{-2}}$ (west arm) \citep{Rodruck_2023}. Kinematic and metallicity maps indicate gas inflows, metal mixing, and disk rebuilding in the northern tail \citep{Buzzo_2021}. A summary of the basic properties is shown in Tabel.~\ref{basics}.

NGC\,1487 provides an ideal case to address key MEDUSA science goals. Its highly disturbed gas kinematics and ongoing starburst activity \citep{Buzzo_2021} create conditions to study magnetic-field structure under strong turbulence and to characterise the spectral behavior of the radio continuum, including energy losses of CREs and free--free absorption in dense star-forming regions. Combined with deep MeerKAT continuum and broadband data, comparing to existing optical, UV, and MUSE observations, this dataset enables a spatially resolved analysis of the interplay between star formation, CRs, and magnetic fields in a low-mass merging system, providing a detailed benchmark for future studies of dwarf galaxies in the MEDUSA survey.

\begin{table}
\centering
\caption{Summary of the basic properties of NGC\,1487.}
\label{basics}
\begin{tabular}{clcccc}
\toprule
&Properties & & &NGC\,1487 &\\
\midrule
&RA (J2000) & & & 03$^{\rm h}$55$^{\rm m}$47.76$^{\rm s}$  &\\
&Dec (J2000) & & & -42$^{\rm d}$21$^{\rm m}$49.139$^{\rm s}$ & \\
&Type & & & Merger& \\
&D / Mpc & & & 12.1 &\\
&i / $^\circ$ & & & 53.4 &\\
&M$_{\rm B}$ / mag  & & & $-17.73$ &\\
&M$_*$ / $10^9$\,M$_\odot$  & & & $4.61\pm 1.24$& \\
&M$_{\rm HI}$ / $10^9$\,M$_\odot$  & & & $1.78$ &\\
&12 + log(O/H)  &  & &  $8.33 \pm 0.16$& \\
&log[H$\alpha$ / (${\rm erg\,s^{-1} cm^{-2}})]$  & & & $-11.63 \pm 0.04$ &\\
&log[SFR / (M$_\odot$yr$^{-1}$)]  &  & & $-1.36$ &\\
&log[$\Sigma_\text{SFR}$ / (M$_\odot$yr$^{-1}$kpc$^{-2}$)]  & & & $-1.93 \pm 0.03$ &\\
&$\varv_{\rm disp}$ / km\,s$^{-1}$  & & & $12.55\pm 2.75$& \\
&$\varv_{\rm rot}$ / km\,s$^{-1}$  & & & $113.7 \pm 3.3$& \\
&$S_{\rm 60 \upmu m}$ / Jy & & & 3.25&\\
&$S_{\rm 100 \upmu m}$ / Jy & & & 6.39&\\
\bottomrule
\end{tabular}
\tablefoot
{The basic properties are coordinates $^{a}$, morphology type $^{b}$, distance $^{c}$, inclination $^{a}$, absolute magnitude in B-band $^{a}$, stellar mass $^{c}$, H{\sc i} mass $^{d}$, metallicity $^{c}$, H$\alpha$ flux $^{e}$, star formation rate $^{e}$, star formation rate surface density $^{e}$, dispersion velocity $^{f}$, rotation velocity corrected for inclination $^{a}$ and far infrared luminosities $^{g}$.
References: $^{(a)}$ LEDA database; 
$^{(b)}$ \citet{Vorontsov_1959}; 
$^{(c)}$ \citet{Buzzo_2021}; 
$^{(d)}$ \citet{Oey_2007}; 
$^{(e)}$ \citet{Meurer_2007}; 
$^{(f)}$ \citet{Mengel_2008};
$^{(g)}$ \citet{IRAS}.
}
\end{table}

\begin{table*}
\centering
\caption{Summary of the radio observations of  NGC\,1487.}
\begin{tabular}{cccccccc}
\toprule
\text{Date} & \text{Obs. Time} & \text{Array} & \text{Central Frequency} & \text{Bandwidth} & \text{Primary} & \text{Phase} & \text{Polarisation Angle} \\ & \text{[min]} & \text{Configuration} & \text{[MHz]} & \text{[MHz]} & \text{Calibrator} & \text{Calibrator}  & \text{Calibrator}\\
\midrule
MeerKAT & & & & & & &\\
12.10.2023 & 180 & -- & 1280 & 856 & J0408-6545 & J0440-4333 & J0521+1638\\
15.10.2023 & 180 & -- & 1280 & 856 & J0408-6545 & J0440-4333 & J0521+1638 \\
\hline
ATCA & & & & & & &\\
\text{05.05.2023} & 240 & 1.5A & 2100 & 2048 & J1934-638 & J0438-436 & J1934-638    \\
\text{06.05.2023} & 300 & 1.5A & 2100 & 2048 & J1934-638 & J0438-436 & J1934-638   \\
\text{21.07.2023} & 270 & 6D & 2100 & 2048 & J1934-638 & J0438-436 & J1934-638   \\
\text{08.08.2023} & 300 & 6D & 2100 & 2048 & J1934-638 & J0438-436 & J1934-638   \\
\text{20.08.2023} & 180 & 6D & 2100 & 2048 & J0823-500 & J0438-436 & J0823-500   \\
\text{10.09.2023} & 630 & H168 & 2100 & 2048 & J1934-638 & J0438-436 & J1934-638   \\
\text{24.03.2024} & 300 & 6A & 5500/9000 & 2048/2048 & J1934-638 & J0438-436 & J1934-638   \\
\text{10.08.2024} & 360 & 1.5A & 5500/9000 & 2048/2048 & J1934-638 & J0438-436  & J1934-638 \\
\text{24.08.2024} & 360 & 1.5A & 5500/9000 & 2048/2048 & J1934-638 & J0438-436 & J1934-638 \\
\text{21.09.2024} & 360 & 1.5A & 5500/9000 & 2048/2048 & J1934-638 & J0438-436 & J1934-638  \\
\text{11.02.2025} & 420 & EW352 & 5500/9000 & 2048/2048 & J1934-638 & J0438-436 & J1934-638  \\
\bottomrule
\end{tabular}
\label{obs}
\end{table*}

\subsection{Observations}
NGC\,1487 has been observed with MeerKAT (project ID: SCI-20230907-ST-01; PI: S.Taziaux) in L-band (1.28\,GHz) and ATCA (project ID: C3531; PI: S.Taziaux) in L/S-band (2.1\,GHz), C-band (5.5\,GHz) and X-band (9\,GHz) between 5-May-2023 and 2-February-2025. The configurations, total observation time, frequency ranges and calibrators used during the observations are summarized in Table~\ref{obs}. The listed observing time represents the total duration, including a 25\,\% overhead. For one ATCA observation, we used the secondary flux calibrator J0823-500 instead of the primary calibrator J1934–638, and the flux scale was transferred via phase bootstrapping\footnote{The corresponding technical keyword in \texttt{miriad} is \texttt{mfboot}.}.

\subsection{Initial calibration}
\label{datareduction}

\subsubsection{MeerKAT observations}
\label{meerkta_obs}

The MeerKAT observations at L-band, covering the frequency range 856 to 1712\,MHz, were recorded using 32768 frequency channels. These data were averaged to 4096 209.984\,kHz-wide channels for calibration at a central frequency of 1284\,MHz. On both days of observations,  J$0408\textrm{--}6545$ was used to set the absolute flux density scale, solve the bandpass, and because it is weakly polarisation, it was also used to solve the on-axis instrumental leakage. The complex gains were solved using an unresolved bright source, J$0440\textrm{--}4333$, and the absolute polarisation was calibrated using 3C\,138. Calibration was performed following standard procedure using the \texttt{CASA package} \citep[][]{casa}, wherein, the absolute flux scale for J$0408\textrm{--}6545$ was set using the task \texttt{setjy} following the strategy provided by the SARAO,\footnote{\url{https://skaafrica.atlassian.net/wiki/spaces/ESDKB/pages/1481408634/Flux+and+bandpass+calibration}} and the polarisation model for 3C\,138 was adopted from \cite{PerleyButler2013}. 

\begin{table*}
\centering
\caption{Summary of the frequency dependent radio polarimetry parameters.}
\begin{tabular}{llcccc}
\toprule
 Parameters  & MeerKAT   & & ATCA &  \\  
 & L-band   & L/S-band & C-band & X-band \\
\midrule
Resolution / $\arcsec$ & $8.56 \times 7.65$ & $5.86 \times 3.02$ & $3.28 \times 1.28$ & $2.64 \times 0.93$ \\
$\sigma_I$ / $\upmu$Jy beam$^{-1}$ & 4.0 & 8.0 & 5.5 & 5.5 \\
\hline
$\sigma_{QU}$ / $\upmu$Jy beam$^{-1}$ & 2.4 & 8.8 & 8.0 & -- \\
$\sigma_{\rm PI}$ / $\upmu$Jy beam$^{-1}$ RMSF$^{-1}$ & 1.54 & 7.67 & 8.39 & -- \\
mean$_{\rm PI}$ / $\upmu$Jy beam$^{-1}$ RMSF$^{-1}$ & 11.8 & 54.1 & 34.6 & -- \\
\hline
Resolution / $\arcsec$ & 22 & 12 & 6 & -- \\
Total images & 43 & 88 & 113 & -- \\
Max. FD / rad m$^{-2}$ & 699 & 1498 & 3992 & -- \\
n$_{\rm channels}$ & 283 & 155 & 219 & -- \\
RMSF FWHM / rad m$^{-2}$ & 49.56 & 97.29 & 183.14 & -- \\
$\lambda_0^2$ / m$^2$ & 0.063 & 0.019 & 0.003 & -- \\
$\nu_{\rm ref}$ / MHz & 1194 & 2166 & 5430 & -- \\
\bottomrule
\end{tabular}
\tablefoot{
The table lists the angular resolution and rms noise in the Stokes\,$I$ and combined Stokes\,$Q/U$ maps. $\sigma_{\rm PI}$ and mean$_{\rm PI}$ denote the rms and mean polarised intensity, respectively. The RM-synthesis parameters include the common angular resolution of the Faraday cube, the number of usable Stokes\,$Q$ and $U$ channel images due to oversampling, the maximum Faraday depth sampled together with the number of frequency channels, the RMSF FWHM, and the weighted mean $\lambda_0^2$ with its corresponding reference frequency $\nu_{\rm ref}$.
}
\label{obs_prop}
\end{table*}

\subsubsection{ATCA observations}
\label{atca_obs}
The ATCA observations covering a frequency range from 1.1--3.1\,GHz (L/S-band), 3.9--7.1\,GHz (C-band) and 8--11\,GHz (X-band). We employed calibration procedures based on the Multichannel Image Reconstruction Imaging Analysis and Display \citep[\texttt{miriad;}][]{miriad}, followed by a standard procedure using \texttt{mfcal} for bandpass calibration, \texttt{gpcal} to derive the gains and instrument leakage. To mitigate the impact of radio frequency interference (RFI) during flux and phase calibration, the edges of the dataset have been flagged using \texttt{uvflag}, we utilised the interactive flagging tool \texttt{bflag}. Additionally, automated flagging routines were applied with the task \texttt{pgflag} to address interference for the source. Any additional corrupted data identified during calibration was manually flagged. 

\subsection{Stokes\,$I$ imaging and self-calibration}
Initially, after cross-calibration, we employed iterative imaging and self-calibration using WSClean \citep{wsclean} and CASA \citep{casa}, (phase-only, frequency-independent, with a solution interval of 3\,min down to 10\,s for MeerKAT observations and 5\,min down to 60\,s for the ATCA observations) until image quality reached convergence with a Briggs weighting with robust = $-1, -0.5, -0.3, 0, 0.3$ to slowly reconstruct the diffuse emission. 
Multifrequency and multiscaling CLEANing \citep{hoegbom_1974} utilising interactive masks using \texttt{flint\_masking}\footnote{\url{https://github.com/flint-crew/flint}} (Galvin in preparation) around visible sources to minimise artefacts and flux scattering and applying detection thresholds to retain genuine emission only.
The continuum images for each frequency were generated using Briggs weighting with $\texttt{robust} = 0$, resulting in resolutions, noise levels, and image sizes, as provided in Table\,\ref{obs_prop}. 
The maximum angular extent of NGC\,1487 is approximately $3\,\arcmin$. With MeerKAT’s shortest baselines of 29\,m ($\approx0.12\,{\rm k}\lambda$) and ATCA’s shortest baselines ranging from 31\,m to 77\,m ($\approx0.2-0.5\,{\rm k}\lambda$), the corresponding largest angular scales are $27.7\,\arcmin$ and $6-16\,\arcmin$, respectively. Since these scales are much larger than the size of NGC\,1487, the combination of short and long baselines ensures that the observations recover all spatial scales of interest, and missing flux is not expected to be an issue.
\begin{figure*}
    \centering
    \includegraphics[width=\linewidth]{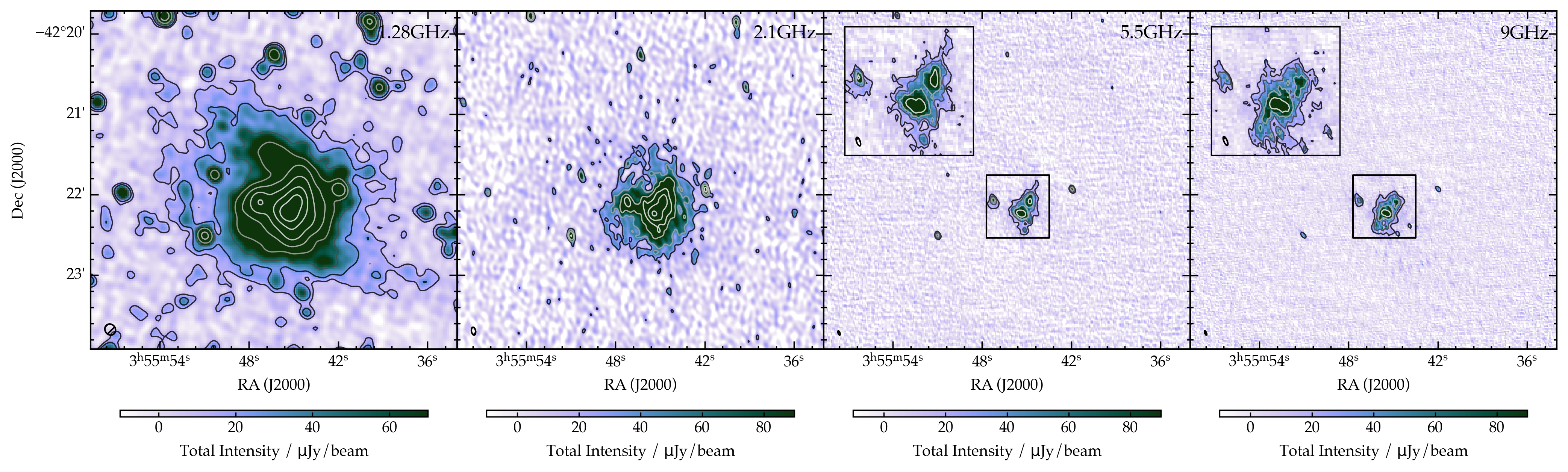}
    \caption{\textit{Left:} Total intensity map at 1.28\,GHz of NGC\,1487 with a noise level $\sigma_I$ of $4\,\upmu$Jy/beam observed with MeerKAT. \textit{Middle left:} Total intensity map at 2.1\,GHz of NGC\,1487 with a noise level $\sigma_I$ of $8\,\upmu$Jy/beam observed with ATCA. \textit{Middle right:} Total intensity map at 5.5\,GHz of NGC\,1487 with a noise level $\sigma_I$ of $5.5\,\upmu$Jy/beam observed with ATCA. \textit{Right:} Total intensity map at 9\,GHz of NGC\,1487 with a noise level $\sigma_I$ of $5.5\,\upmu$Jy/beam observed with ATCA. The overlaid contours of the total intensity start at $3\,\sigma$ and increase by factor of 2. The resolution of the different maps can be taken from Table~\ref{obs_prop} and is shown in the bottom left corner.}
    \label{stokesi}
\end{figure*}

\subsection{Polarisation imaging}
For polarisation imaging, we individually imaged the Stokes\,$Q$ and $U$ parameters with a channel width of 13.5\,MHz to mitigate bandwidth depolarisation effects at a Briggs robust weighting of $0.5$ to trace the diffuse polarised emission. We used a higher positive robust weighting parameter for the Stokes\,$Q$ and Stokes\,$U$ imaging than for Stokes\,$I$ in order to enhance the sensitivity to diffuse polarised emission. 
We use some additional parameters during polarisation imaging with \textsc{WSclean}, such as \texttt{-squared-channel-joining}, \texttt{-join-polarizations} and \texttt{-squared-channel-joining} to improve the quality of the reconstructed polarised signal by reducing leakage between Stokes parameters, ensuring a more consistent spectral response across channels, and suppressing deconvolution artefacts in low–signal-to-noise regions. This results in a more reliable recovery of faint, extended polarised structures.
The polarised intensity ($PI$) was derived as the maximum value along the Faraday axis of the polarised intensity cube, calculated from $PI = \sqrt{Q^2 + U^2}$ \citep[][]{Wardle_1974}.

\subsection{Rotation measure synthesis}
The resulting images have been primary beam corrected and convolved to lowest common resolution to perform rotation measure (RM) synthesis \citep[][]{Burn_1966, Brentjens_2005, Heald_2009_II} using the \textsc{RMtools}\footnote{\url{https://github.com/CIRADA-Tools/RM-Tools}} \citep[][]{rmtools} suite task \texttt{rmsynth3d}. The parameters, which depends on the measurement set, such as resulting images, common resolution, and sampling and stepsize numbers are summarised in Table~\ref{obs_prop}. 

To deconvolve the rotation measure spread function (RMSF) from the Faraday dispersion function (FDF) and thereby suppress sidelobes and artificial structure, we performed RM-clean using the \textsc{RMtools} task \texttt{rmclean3d}. RM-clean requires an estimate of the noise in the complex polarisation signal, so we cleaned down to a threshold of $5\sigma_{QU}$ \citep{Heald_2009_II}. Here, $\sigma_{QU}$ is not the uncertainty on the polarised intensity; rather, it represents the effective noise level in the complex $Q+iU$ plane. We derived $\sigma_{QU}$ from independent noise estimates in Stokes\,$Q$ and $U$ ($\sigma_Q$ and $\sigma_U$), measured from regions at Faraday depths where no signal is present. The combined noise level was then computed using standard error propagation for the complex quantity $Q+iU$: $\sigma_{QU} = \sqrt{\frac{Q^2}{Q^2+U^2}\sigma_Q^2+\frac{U^2}{Q^2+U^2}\sigma_U^2}$.

After all, we perform the \texttt{rmtools\_peakfitcube} task, taking a 3D Faraday depth cube, analysing the RM spectrum at each spatial pixel and fitting the main peaks. 
The polarised intensity is corrected for Ricean bias\footnote{Peak polarised intensity after Ricean-bias correction, applied when $\mathrm{snrPIfit} > 5$, using $PI_{\rm eff} = \sqrt{PI^{2} - 2.3\,\sigma_{\rm FDF}^{2}}$}. 
The polarisation angle $\chi$ was determined from the Stokes parameters using $\chi = \frac{1}{2} \arctan\left(\frac{U}{Q}\right)$, providing the orientation of the electric field vectors in the plane of the sky, using the Stokes\,$Q$ and $U$ maps at the reference frequency $\nu_{\rm ref}$ listed in Table~\ref{obs_prop}.  The intrinsic magnetic field orientation was then obtained by derotating the observed polarisation angles using $\chi_0 = \chi - \text{RM} \, \lambda_0^2 + \frac{\pi}{2}$, where $\lambda_0^2$ is the weighted mean of the observed $\lambda^2$ values. All the relevant parameter values are presented in Table~\ref{obs_prop}.

\section{Properties of radio continuum}
\label{radiocont}
The radio spectrum of a galaxy is typically a superposition of free--free and synchrotron emission \citep[e.g.][]{klein_2018}, both of which can be written as a power-law in the optical thin regime. At frequencies below 10\,GHz, the synchrotron dominates with a spectral index ($S\propto\nu^{\alpha_{\rm nth}}$) of approximately $-0.7$ \citep{Basu_2015, Werhahn_2021}. By isolating synchrotron emission from free-free emission, the total magnetic field strength can be determined based on the assumption of energy equipartition between CRs and magnetic fields (see Sect.~\ref{equipartition_sect}). Because free-free emission does not exhibit polarisation, while synchrotron emission does, the presence of polarised emission serves as a clear indicator of synchrotron emission, providing crucial information about the magnetic field component in the plane of the sky (see Sect.~\ref{pol}).

\subsection{Total intensity}

Fig.~\ref{stokesi} shows the radio continuum emission in NGC\,1487. At 1.28\,GHz, the diffuse emission extends up to the base of the tidal arms, which are also visible in the H$\alpha$ and GALEX-FUV images (see Fig.~\ref{optical}). At higher frequencies (up to 9\,GHz), the brightest compact regions, previously identified by \citet{Aguero_1997} and \citet{Buzzo_2021}, remain detectable. These regions are predominantly ionised by ongoing star formation and exhibit relatively low dust content. A comparison between the radio continuum and H$\alpha$ emission shows a strong spatial correlation, with emission peaks closely coinciding; the two most prominent radio-emitting regions correspond to H{\sc ii} regions. The radio continuum traces recent star formation, with low-frequency emission dominated by synchrotron radiation from CREs and higher-frequency emission increasingly contributed by free--free emission. The combined emission provides a complementary, dust-unbiased probe of star formation activity.

\subsection{Spectral analysis}
\subsubsection{Full integrated spectral energy distribution}
\label{sed_sec}
\begin{figure}
    \centering
    \includegraphics[width=\linewidth]{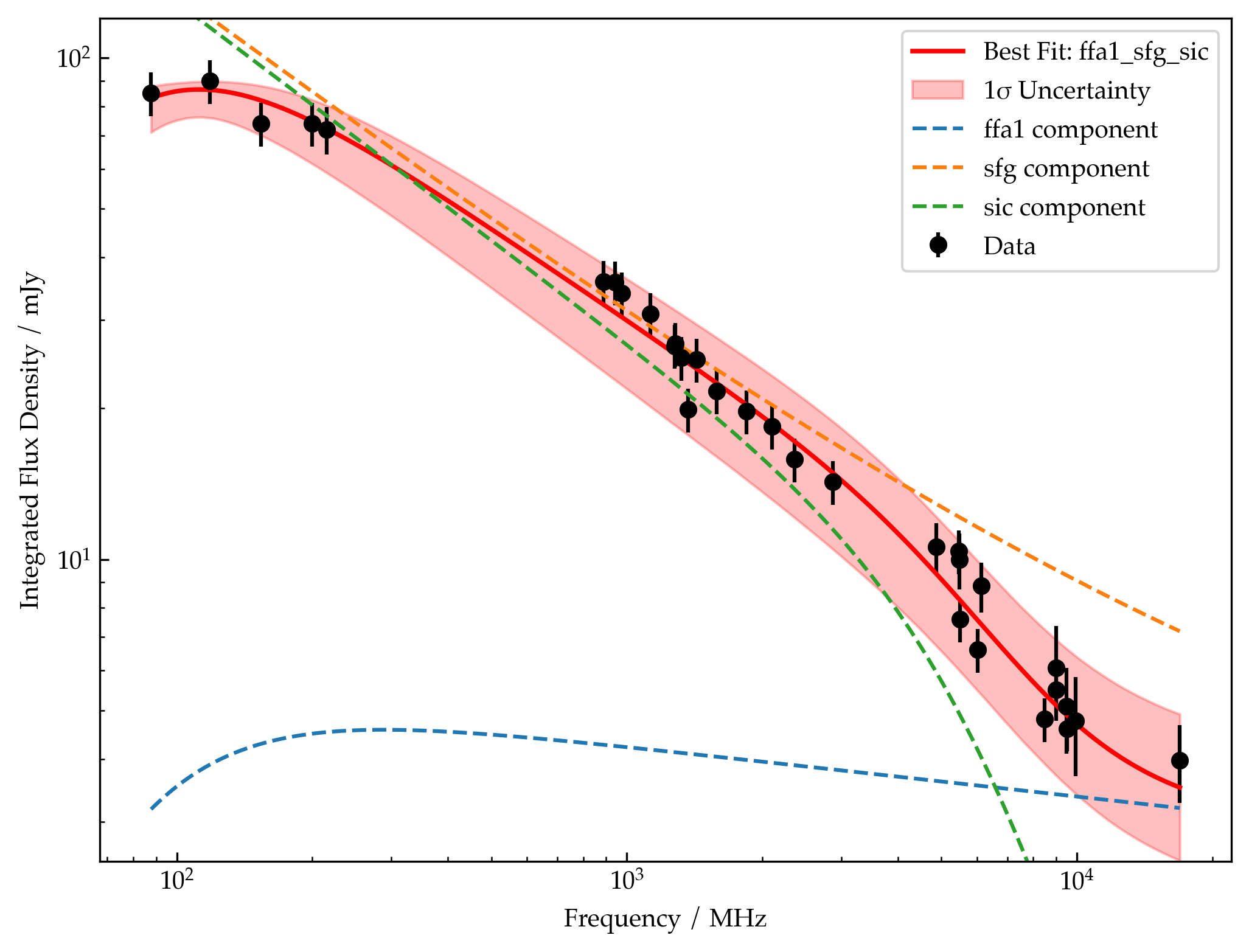}
    \caption{Spectral energy distribution of NGC\,1487. The plot displays total intensity data with different observations: GLEAM-X data \citep{gleamx}, RACS-Low data \citep{racs}, ASKAP EMU main survey \citep{emu_main}, MeerKAT 1.28\,GHz data, ATCA L/S-band, C-band, and X-band data presented in this work, and further data from the work of \citet{Grundy_2025}. We show the best-fit in red with $1\sigma$ uncertainty sampled by \texttt{EMCEE}. The blue, orange, and green dotted lines represent the different components of the best-fit model.}
    \label{sed}
\end{figure}

\begin{figure*}
    \centering
    \includegraphics[width=\linewidth]{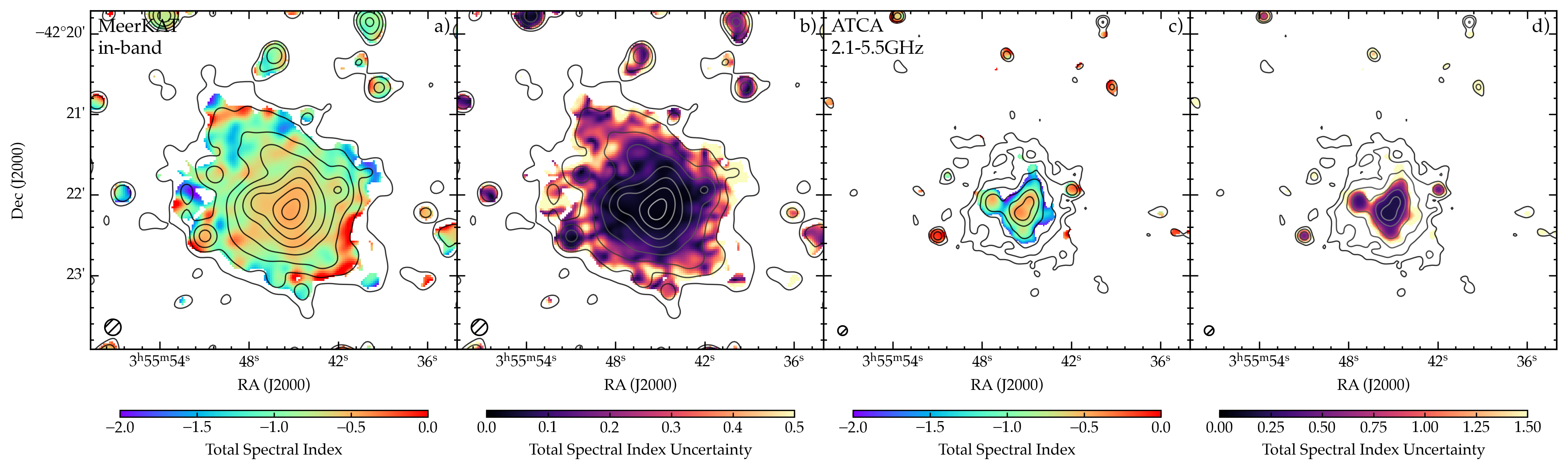}
    \caption{Inband total spectral index of the MeerKAT L-band (\textit{panel a)}) and its uncertainty (\textit{panel b)}) between the frequencies 0.96 and 1.59\,GHz overlaid with radio emission at frequency of 1.28\,GHz starting at $3\sigma$ and increasing by a factor of 2 ($\sigma=9\,\upmu$Jy/beam). The resolution of $12\arcsec$ is shown in the bottom left corner.
    Two-point total spectral index (\textit{panel c)}) and its uncertainty (\textit{panel d)}) between ATCA L/S-band (2.1\,GHz) and C-band (5.5\,GHz) overlaid with radio emission at frequency of 2.1\,GHz starting at $3\sigma$ and increasing by a factor of 2 ($\sigma=11.3\,\upmu$Jy/beam). The resolution of $7\arcsec$ is shown in the bottom left corner.
    }
    \label{spix}
\end{figure*}

To cover a broad range of the radio continuum emission, we use the central frequencies 1.28\,GHz, 2.1\,GHz, 5.5\,GHz and 9\,GHz and additionally, we split the MeerKAT data set into six 143\,MHz wide sub-bands, ATCA L/S-band into six 341\,MHz wide sub-bands, and C- and X-band into each two sub-bands of 683\,MHz width. We applied a $3\,\sigma$ clipping threshold and computed the integrated flux density, summing all emission over its resolved size using \texttt{RadioFluxTools}\footnote{\url{https://gitlab.com/Sunmish/radiofluxtools}}, across the individual frequency slices. To ensure the reliability of the measurements, we have to discard one of the MeerKAT slices and one of the ATCA C-band slices, as the quality is too low. 
But we also incorporate literature data from the GaLactic and Extragalactic All-sky MWA survey eXtended (GLEAM-X) data \citep[][]{gleamx}, which cover frequencies from 87\,MHz to 221\,MHz. We also include the Australian SKA Pathfinder (ASKAP) survey, Rapid ASKAP Continuum Survey (RACS-Low) data \citep[][]{racs} at 885\,MHz. Also, data from the ASKAP Evolutionary Map of the Universe (EMU) main survey \citep{emu_main} at 943\,MHz are used in this study.
Furthermore, we use literature data from \citet{Grundy_2025}. The full values of the flux densities can be found in Appendix~\ref{flux_density_table_sect} in Table~\ref{flux_density_table}

To study the spectral behavior, we fit several models to the spectral energy distribution to gain insight into the underlying physical processes. For this, we fit different models, explained in \citet{galvin_characterizing_2016}, \citet{Grundy_2025} and \citep{Taziaux_2025_chilling} to the SED, and we find that a mixture of free--free absorption (\texttt{ffa1}) at low frequency, a superposition of free--free emission and synchrotron emission at intermediate frequency (\texttt{sfg}), and a break due to inverse Compton losses at higher frequencies (\texttt{sic}) fits the data best. We call this model \texttt{ffa1\_sfg\_sic}.
The flux density is given by,
\begin{equation}
S_{\nu} = (1 - e^{-\tau}) \left( B + A \left( \frac{\nu}{\nu_0} \right)^{0.1 + \alpha_{\rm nth}} \frac{1}{1 + \left( \frac{\nu}{\nu_b} \right)^{\Delta \alpha}} \right) \left( \frac{\nu}{\nu_0} \right)^2\,,
\end{equation}
where $A$ represents the synchrotron emission, $\alpha_{\rm nth}$ is the non-thermal spectral index, and $B$ is the free--free emission at a given frequency $\nu_0$. $\nu_{\rm b}$ is the break frequency in the spectrum, and $\Delta \alpha$ represents the change in the non-thermal spectral index due to synchrotron and inverse Compton losses, assuming continuous electron injection from massive star formation. 

At low radio frequencies, synchrotron emission can be attenuated by free-free absorption. This occurs when the synchrotron-emitting regions are embedded within or behind optically thick ionised gas, which absorbs some of the radiation and produces a spectral turnover or flattening, often observed in star-forming galaxies \citep[e.g.][]{Gajovic_2025, Gajovic_2024, Grundy_2025, taziaux_2025}. The strength of this absorption depends on the density and distribution of the ionised gas, and the optical depth $\tau$ reaches unity at a characteristic `turnover' frequency. At high frequencies, the spectrum may exhibit further steepening due to energy losses of CREs, primarily via synchrotron radiation and inverse Compton scattering with cosmic microwave background. This results in a gradual steepening of the synchrotron spectral index above a characteristic `break' frequency $\nu_{\rm b}$.

In Fig.~\ref{sed}, we show the spectral energy distribution of NGC\,1487 using a non-linear least-squares method (\texttt{leastsq}) with six free parameters in the best-fitting model (\texttt{lffa1\_sfg\_sic}). The fit yields a reduced chi-square of $\chi^2_{\rm red} = 1.10$, indicating a statistically acceptable match to the data. Parameter uncertainties were estimated from the fit itself and supplemented with Monte-Carlo sampling to ensure physically meaningful variations (e.g. positive amplitudes and characteristic frequencies).  The non-thermal spectral index is $\alpha_{\rm nth} = -0.678 \pm 0.085$, consistent with a moderately steep spectrum indicative of aged CREs. The spectral break occurs at $\nu_{\rm b} \approx 6.2 \pm 1.3$\,GHz, marking the frequency where synchrotron and inverse Compton losses begin to steepen the spectrum. All the parameter values and its uncertainties are shown in Table.~\ref{sed_fit_params} in Appendix~\ref{flux_density_table_sect}.

\subsubsection{Pixel-based spectral index}
\label{spix_pixel}
Spectral indices derived from observed total intensity radio continuum emission maps serve as crucial information for explaining the underlying emission mechanisms, and energy loss/gain processes of the synchrotron emitting CREs spatially. We present in Fig.~\ref{spix}, the total spectral index maps of NGC\,1487. 
To minimise additional uncertainties, we do not apply a thermal correction to the maps.

To obtain the spatially resolved in-band spectral index, we fit a single power-law spectrum to the flux density in each pixel across the available sub-bands. Panels a) and b) of Fig.~\ref{spix} show the resulting MeerKAT in-band spectral index map and its associated uncertainty. For this purpose, we split the measurement set into six sub-bands (0.96, 1.09, 1.22, 1.34, 1.47, and 1.59\,GHz), convolved each sub-band image to a common resolution of $12\arcsec$, clipped each map at a noise level of $3\sigma$, and then performed a linear fit in $\log S \textrm{--}\log \nu$ space for every pixel.
In the center of the dwarf galaxy, we find spectral index values reaching up to $-0.45\pm0.02$, and it seems that overall they steepen radially toward the edges of the galaxy. But, the map reveals a clear substructure beyond a simple monotonic radial trend. Several compact regions within the disk exhibit flatter spectra compared to their surroundings, most notably toward the western and northeastern parts of the system, following the emission from the H$\alpha$ map (see middle panel of Fig.~\ref{optical}), which could trace sites of recent CRE injection, similar to IC\,10 \citep{Basu_2017}. In contrast, more extended patches of steep spectral index ($\alpha \lesssim -1$) are present toward the southern and northern periphery, suggesting enhanced energy losses or efficient transport of CREs out of the star-forming disk. The spectral index distribution is therefore not azimuthally symmetric due to a combination of spatially intermittent regions of CRE injection, non-uniform CRE transport process\footnote{Non-uniform CRE transport denotes possible spatial variations in diffusion/advection and energy-loss conditions that cause CREs to propagate differently across the galaxy.}, and possibly localised outflows.
Note that at the edges of the galaxy, the total spectral index value flattens significantly and even reaches positive values (e.g. $0.37\pm0.82$ in the southeast side of the galaxy), but these are likely artefacts caused by low signal-to-noise. 

In the panel c) and d) of Fig.~\ref{spix}, we show the two-point spectral index and its uncertainty map between the ATCA L/S-band (2.1\,GHz) and C-band (5.5\,GHz). Both maps were convolved to a common resolution of $7\arcsec$ and clipped at the $3\sigma$ level. The spectrum across the two main cores is relatively flat, with values of $-0.36 \pm 0.18$ and $-0.42 \pm 0.19$, and slightly steeper in the eastern core with $-0.46\pm0.29$. The flat spectrum closely follows the distribution of H{\sc ii} regions seen in the H$\alpha$ map. We note that the core spectral indices differ between the MeerKAT inband and the ATCA L/S– and C-band measurements (panel a) and c) of Fig.~\ref{spix}), mainly because no thermal correction has been applied. The thermal fraction at 1.28\,GHz is almost negligible, while it becomes more significant at 2.1\,GHz and even more at 5.5\,GHz \citep[e.g.][]{Parnovsky_2018, hindson_radio_2018}. Consequently, the core appears flatter in the ATCA two-point spectral index map due to the contribution of thermal emission at these higher frequencies. Outside the star-forming regions, the spectrum declines rapidly, with very steep values measured just beyond the knots. This steepening suggests strong energy losses of the CREs, likely dominated by inverse Compton cooling or by transport through outflows or galactic winds.

\begin{figure*}
    \centering
    \includegraphics[width=\linewidth]{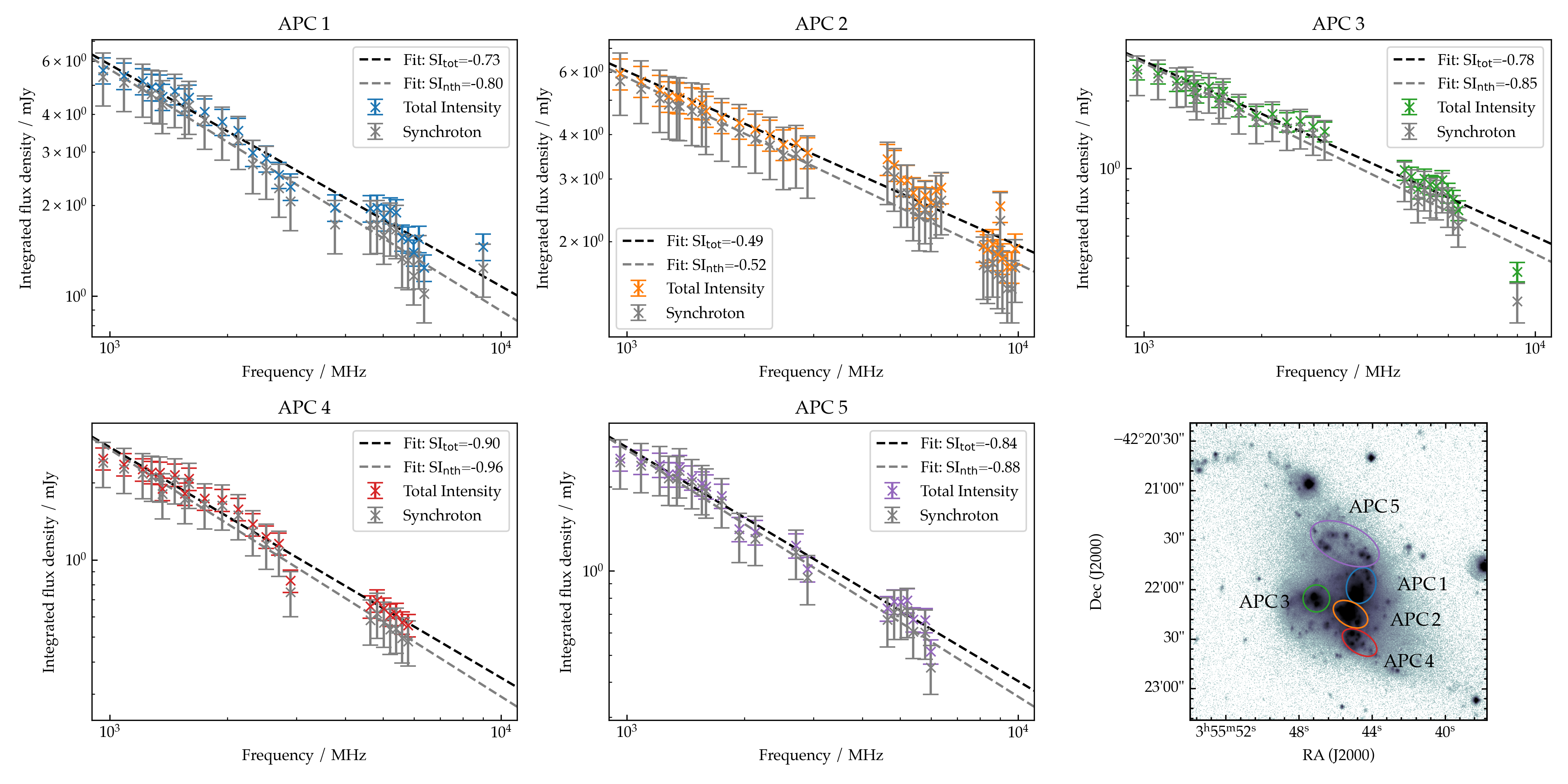}
    \caption{Region dependent spectral energy distribution, showing total and thermal corrected flux density of MeerKAT L-band, ATCA L/S-, C- and X-band data. The bottom right corner panel shows the H$\alpha$ map with the individual regions overlaid in different colors.}
    \label{region_sed}
\end{figure*}

\subsection{Region dependent spectral analysis}
\label{region_sed_sect}
The GALEX–FUV and H$\alpha$ maps in Fig.~\ref{optical} reveal that NGC\,1487 consists of several distinct star-forming components, which can be analysed individually.
Following the classification of \citet{Buzzo_2021}, we divide the system into five regions: APC\,1, APC\,2, and APC\,3 correspond to the three main central star-forming knots, while APC\,4 and APC\,5 trace the base of the southern and northern tidal features, respectively. Fig.~\ref{region_sed} shows SEDs for these regions and an H$\alpha$ image overlaid with the location of the regions. Each dataset was divided into eight sub-bands, and the flux density was measured for each region. Owing to the low signal-to-noise ratio in the X-band, we only show the sub-band measurements for the main core region (APC\,2) and the integrated full band measurement for APC\,1 and APC\,3. For the regions tracing the base of the tidal features, the ATCA X-band data are excluded, as only noise was detected.

To get the synchrotron emission from the total emission we have to apply a thermal correction. Additionally, as starburst dwarf galaxies are generally optically thick  \citep{kepley_role_2010, Basu_2017, taziaux_2025}, we also have to account for absorption. It is possible that free--free absorption in a thermal plasma coexists with the relativistic electrons responsible for the synchrotron emission. Therefore, we take internal absorption into account \citep{Tingay_2003}. Normally, we do not have absorption at these high frequencies, but we still include it in the measurement. We ignore external absorption as we do not expect free--free absorption by a screen of ionised gas located outside the synchrotron emitting region. The thermal correction can be formulated as,
\begin{equation}
    S_{\rm th}(\nu) = S_0(\nu)\,\nu^{-0.1} \biggl(\frac{1-{\rm e}^{-\tau_{\rm ff}}}{\tau_{\rm ff}}\biggr) ,
    \label{thcorr}
\end{equation}
depending on the free--free optical depth $\tau_{\rm ff}$, the thermal emission $S_0(\nu)$  which can be obtained by the free--free emission density function in relation to the H${\alpha}$ flux \citep[][Eq.\,3]{Deeg_1997} and the frequency $\nu$. The full calculation of the thermal correction is detailed in the Appendix~\ref{thermalcorrection}, where we also show that the free--free optical depth $\tau_{\rm ff}$ depends on the electron temperature $T_{\rm e}$ (see Eq.~\ref{opticaldepth} in Appendix~\ref{thermalcorrection}). 
We assume a 10\,\% uncertainty for the total intensity. For the thermal-corrected intensity, we adopt a 20\,\% uncertainty to account for additional factors, including variations in the electron temperature ($T_{\rm e} = 10000 \pm 3000$\,K) and uncertainties in the [N{\sc ii}] line flux correction relative to H$\alpha$ measurements \citep{Vargas_2018}. 

For APC\,2, the most actively star-forming region \citep{Buzzo_2021}, we find a flat spectrum. After removing the thermal contribution, the non-thermal index remains flat ($\alpha_{\rm tot}=-0.49 \pm 0.02$; $\alpha_{\rm nth}=-0.52 \pm 0.02$), which is close to the expected injection index of $\alpha_{\rm inj}\approx -0.5$ from diffusive shock acceleration in supernova remnants.
In contrast, APC\,1 and APC\,3 exhibit significantly steeper spectra, with total spectral indices of $\alpha_{\rm tot}=-0.73\pm0.03$ and $\alpha_{\rm tot}=-0.78\pm0.03$, and non-thermal values of $\alpha_{\rm nth}=-0.80\pm0.03$ and $\alpha_{\rm nth}=-0.85\pm0.04$, respectively. The steeper spectra in these regions indicate that synchrotron emission dominates and that the CRE population has already experienced radiative losses. This likely means that APC\,1 and APC\,3 are slightly older star-forming regions compared to APC\,2, with fewer newly accelerated electrons.
In the regions at the base of the tidal arms, APC\,4 and APC\,5, we measure steep spectra with total spectral indices of $\alpha_{\rm tot}=-0.90\pm0.03$ and $\alpha_{\rm tot}=-0.84\pm0.03$, respectively. After correcting for thermal emission, these values become $\alpha_{\rm nth}=-0.96\pm0.03$ and $\alpha_{\rm nth}=-0.88\pm0.03$. Interestingly, \citet{Buzzo_2021} report nearly identical star formation rates for APC\,3 and APC\,4 ($0.04\pm0.01 \, {\rm M_\odot \, yr^{-1}}$ and $0.03\pm0.01 \, {\rm M_\odot \, yr^{-1}}$, respectively).

\subsection{Equipartition magnetic field strength}
\label{equipartition_sect}
A common approach to estimate the magnetic field strength is to assume equipartition between the energy densities of CRs and the magnetic field, although these assumptions carry significant uncertainties \citep{Pfrommer_2004MNRAS, kepley_role_2010, Seta_2019, Ruszkowski_2023, Dacunha_2024}. By assuming that the thermal/non-thermal decomposition is valid and that equipartition can be applied to starburst dwarf galaxies \citep{Chiu_2025}, the magnetic field under equipartition can be determined following \citet{Beck_2005}
\begin{equation}
    B_\text{eq}=\Biggl(\frac{4\pi (2\alpha_{\rm nth}-1)(K_0+1)S_\nu E_\text{p}^{1+2\alpha_{\rm nth}}(\frac{\nu}{2c_1})^{-\alpha_{\rm nth}}}{(2\alpha_{\rm nth} +1)c_2(\alpha_{\rm nth})lc_4(i)}\Biggr)^\frac{1}{(3 - \alpha_{\rm nth})},
    \label{equipartition_eq}
\end{equation}
where $\alpha_{\rm nth}$ is the non-thermal spectral index and $S_\nu$ is the synchrotron intensity at the frequency $\nu$. We adopt a proton-to-electron energy density ratio of $K_0 = 100$, in line with values used in previous studies \citep[e.g.][]{chyzy_magnetized_2016, Basu_2017, heesen_nearby_2022, Stein_2023, taziaux_2025}. The galaxy inclination is taken as $i = 53.4^\circ$, and we assume a path length $l = 5.5\,$kpc, corresponding to the spatial extent over which CREs can be traced. Estimating the path length $l$ is generally challenging due to the uncertain geometry of the galaxy. A detailed description of the constants $c_1$, $c_2$, and $c_4$, as well as the full equipartition magnetic field equation, can be found in \citet{Beck_2005}. It should be noted that the equipartition formula becomes undefined for simple power-law energy spectra for $\alpha_{\rm nth} \ge -0.5$.
We use the MeerKAT L-band total intensity map together with the in-band total spectral index map. At 1.28\,GHz, the thermal fraction is assumed to be negligible, so no thermal correction is applied. 

\begin{figure}
    \centering
    \includegraphics[width=\linewidth]{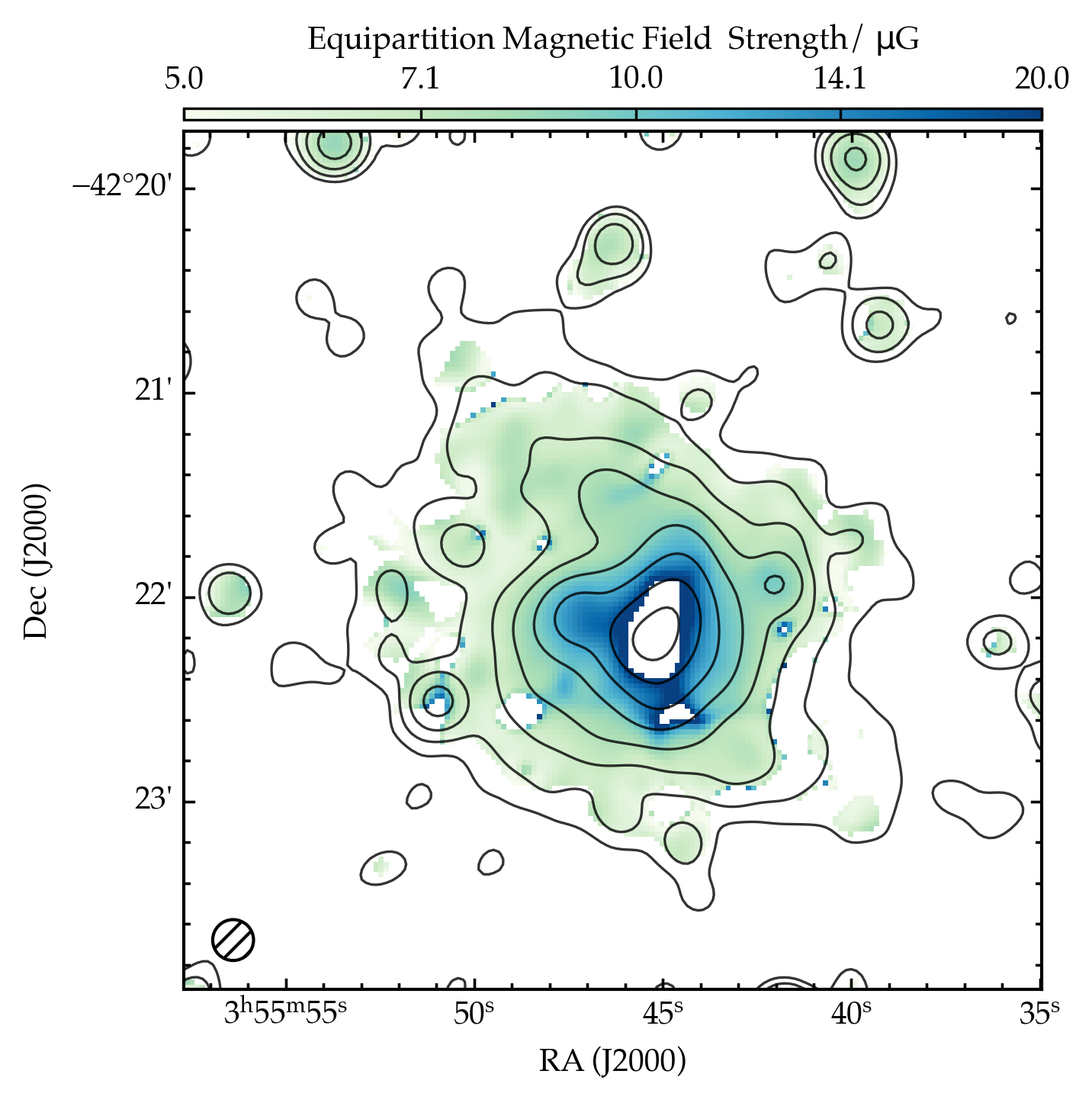}
    \caption{Total magnetic field strength of NGC\,1487 estimated assuming energy equipartition between CR particles and magnetic field. The inner white regions are blanked where the equipartition condition is invalid $(\alpha > -0.5)$. The overlaid contours are the total intensity emission at 1.28\,GHz.}
    \label{equipartition}
\end{figure}

Fig.~\ref{equipartition} shows the equipartition magnetic field strength with strong magnetic field in the center core with values higher than $12\,\upmu$G. The strength decreases radially and takes values at the edges of nearly $5\,\upmu$G. These findings are similar to other dwarf galaxies \citep[e.g.][]{kepley_role_2010, chyzy_magnetized_2016, Basu_2017, heesen_exploring_2018, taziaux_2025}. 

To calculate the magnetic field strength using the equipartition assumption for the different regions, we use the non-thermal spectral index value for each region and calculate the mean flux value on the MeerKAT L-band map. We still assume the same path length of 5.5\,kpc as the physics of the CREs do not change between a small region and the full galaxy. For the more star-forming regions APC\,1, APC\,2 and APC\,3, we calculate a magnetic field strength of $12.67\,\upmu$G, $18.97\,\upmu$G, and $11.13\,\upmu$G, respectively. For the more distant region, APC\,4 and APC\,5, we calculate a magnetic field strength of $10.12\,\upmu$G and $5.46\,\upmu$G.

\section{Magnetic field configuration}
In the following section, the band names refer to the $\nu_{\rm ref}$ values given in Table~\ref{obs_prop}.
\label{mfconfig}
\subsection{Polarised emission}
\begin{figure*}
    \centering
    \includegraphics[width=\linewidth]{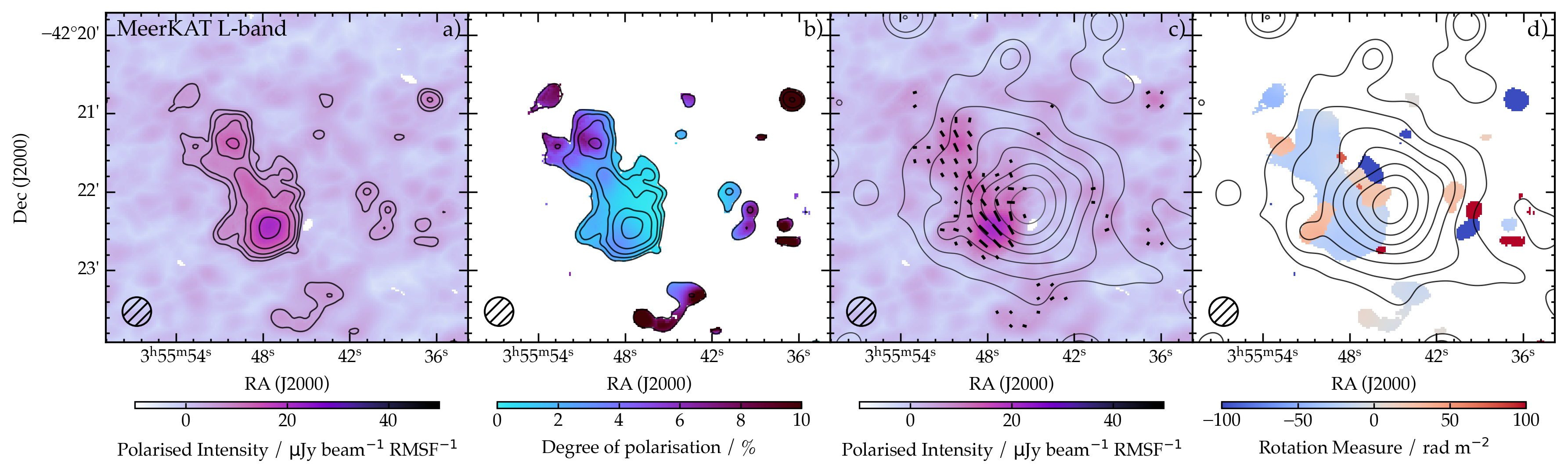}
    \caption{\textit{Panel a):} Polarised intensity at 1194\,MHz with contours starting at $3\sigma_{\rm PI}$ and increase by $\sqrt{2}$ ($\sigma_{\rm PI}$ =$1.54\,\upmu$Jy\,beam$^{-1}$\,RMSF$^{-1}$).
    \textit{Panel b):} Fractional polarisation at 1194\,MHz with PI contours as in panel a).
    \textit{Panel c):} Polarised intensity with total intensity contours at 1194\,MHz starting at $3\sigma$ and increasing by a factor of two ($\sigma = 30\,\upmu$Jy/beam). The  magnetic field orientations are shown in black and scale with the polarised intensity.
    \textit{Panel d):} Foreground-corrected rotation measure with total intensity contours as in panel c). For each panel, the circular beam of $22\arcsec$ appears in the lower left corner.}
    \label{pol_meerkat}
\end{figure*}

\begin{figure*}
    \centering
    \includegraphics[width=\linewidth]{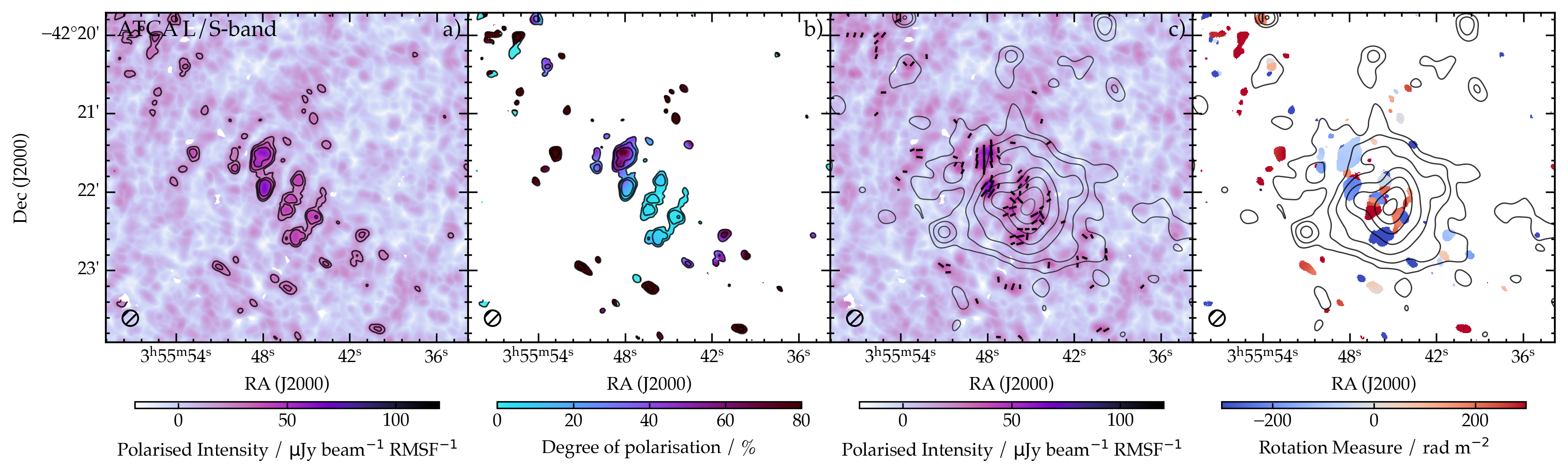}
    \caption{\textit{Panel a):} Polarised intensity at 2166\,MHz with contours starting at $3\sigma_{\rm PI}$ and increase by $\sqrt{2}$ ($\sigma_{\rm PI}$ =$7.67\,\upmu$Jy\,beam$^{-1}$\,RMSF$^{-1}$). 
    \textit{Panel b):} Fractional polarisation at 2166\,MHz with PI contours as in panel a).
    \textit{Panel c):} Polarised intensity with total intensity contours at 2166\,MHz starting at $3\sigma$ and increasing by a factor of two ($\sigma = 15\,\upmu$Jy/beam). The  magnetic field orientations are shown in black and scale with the polarised intensity.
    \textit{Panel d):} Foreground-corrected rotation measure with total intensity contours as in panel c). For each panel, the circular beam of $12\arcsec$ appears in the lower left corner.}
    \label{pol_atca21}
\end{figure*}

\begin{figure*}
    \centering
    \includegraphics[width=\linewidth]{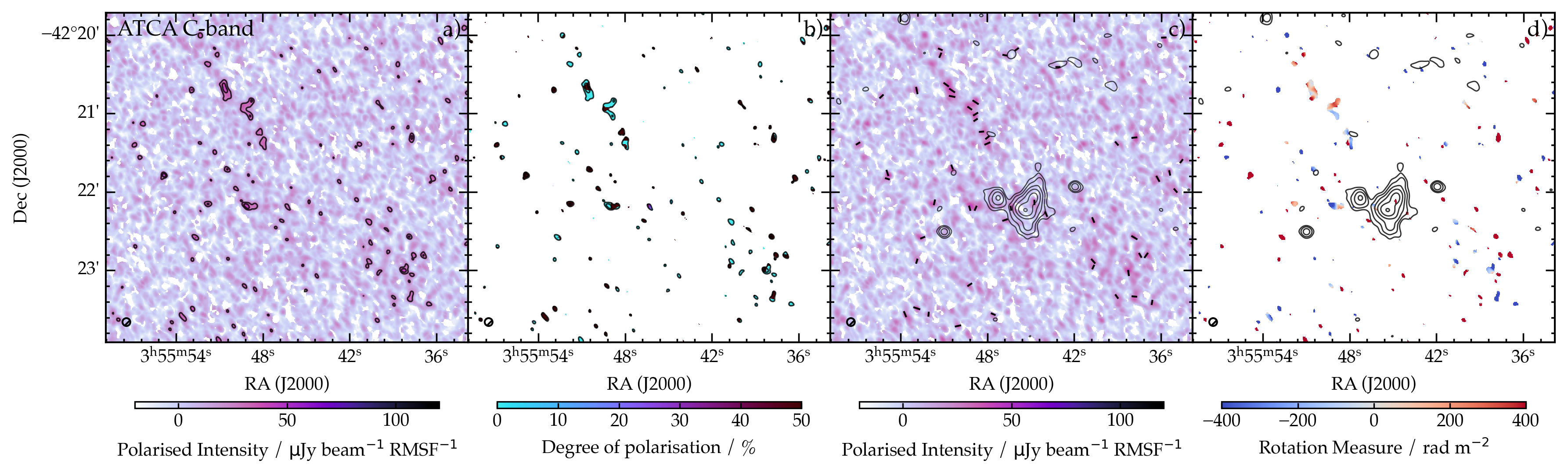}
    \caption{\textit{Panel a):} Polarised intensity at 5430\,MHz with contours starting at $3\sigma_{\rm PI}$ and increase by $\sqrt{2}$ ($\sigma_{\rm PI}$ =$8.39\,\upmu$Jy\,beam$^{-1}$\,RMSF$^{-1}$).
    \textit{Panel b):} Fractional polarisation at 5430\,MHz with PI contours as in panel a).
    \textit{Panel c):} Polarised intensity with total intensity contours at 5430\,MHz starting at $3\sigma$ and increasing by a factor of two ($\sigma = 11\,\upmu$Jy/beam). The  magnetic field orientations are shown in black and scale with the polarised intensity.
    \textit{Panel d):} Foreground-corrected rotation measure with total intensity contours as in panel c). For each panel, the circular beam of $6\arcsec$ appears in the lower left corner.}
    \label{pol_atca55}
\end{figure*}

In Fig.~\ref{pol_meerkat} (MeerKAT L-Band), \ref{pol_atca21} (ATCA L/S-band), and \ref{pol_atca55} (ATCA C-band), we present the polarised intensity, fractional polarisation, magnetic field orientations\footnote{Throughout this work, `magnetic field orientations' refers to Faraday rotation corrected magnetic field orientations.}, and the rotation measure of NGC\,1487. 
To mitigate the strong bias introduced by the positive-definite background, we subtract the background mean (see Table~\ref{obs_prop}) from the polarised intensity map \citep{Heald_2009_II}. To ensure that only significant emission is considered, we apply a clipping threshold of $3\,\sigma_{\rm PI}$. The noise levels for the polarised intensity $\sigma_{\rm PI}$ were estimated from the bias-corrected polarised intensity map, well separated from the phase centre. 
As the effective frequency of the polarisation emission map, generated by RM synthesis, is not at the central frequency but at the weighted mean $\lambda^2_0$, we normalise the Stokes\,$I$ to a reference frequency (see Table~\ref{obs_prop}) to generate the fractional polarisation maps /panel b) of Fig.~\ref{pol_meerkat} -- \ref{pol_atca55}). 

In the MeerKAT L-band (Fig.~\ref{pol_meerkat}), we detect polarised emission in the eastern region of NGC\,1487. This polarised emission extends into the continuum disk. Additionally, we observe regions of high polarised intensity below APC\,3. The star-forming regions show a relatively low fraction of polarisation (panel b), Fig.~\ref{pol_meerkat}), which increases toward the galaxy’s outskirts, reaching values of $6-10\,\%$. In panel c), we show the magnetic field orientation. 
At two locations, we also detect extensions where the magnetic-field orientations appear to reach toward the galaxy’s outskirts.

Fig.~\ref{pol_atca21} shows the polarisation properties of the ATCA L/S-band data. We detect polarised emission between the star-forming regions APC\,1 and APC\,2, as well as around APC\,3, which corresponds to the merger. High polarised emission is also observed above APC\,3, extending nearly to the outskirts of the galaxy and the beginning of the northern tidal arm. In contrast, lower polarised emission is detected at the start of the southern tidal arm. In panel b) of Fig.~\ref{pol_atca21}, the degree of polarisation is low around the star-forming knots, but increases rapidly toward the northern part of the galaxy, reaching nearly 60\,\%. Panel c) shows the magnetic field orientation overlaid on the total intensity. In the northern region, the magnetic field orientations have a similar orientation to those seen in the MeerKAT L-band data and generally point vertically toward the tidal arms. In the central star-forming region, the magnetic field appears chaotic, with orientations pointing in multiple directions and no clear large-scale alignment. This suggests a turbulent magnetic field in the core, influenced by intense star formation and associated gas motions.

Fig.~\ref{pol_atca55} shows the polarisation properties of the ATCA C-band data. Overall, we do not detect significant polarised emission, and the signal is close to the noise level. Only a few small regions show weak polarised emission, primarily along the northern tidal arm. A small detection is also observed on the eastern side of the star-forming region APC\,3.
For the ATCA X-band data, no polarised emission is detected, so it is not shown here.
Although depolarisation is reduced at higher frequencies and one would therefore expect stronger polarised emission, the synchrotron surface brightness, and consequently the polarised intensity, declines steeply with frequency. In our ATCA X-band data, the resulting polarised intensity falls below the noise level, and no significant polarised emission is detected.

\begin{figure*}
    \centering
    \includegraphics[width=0.85\linewidth]{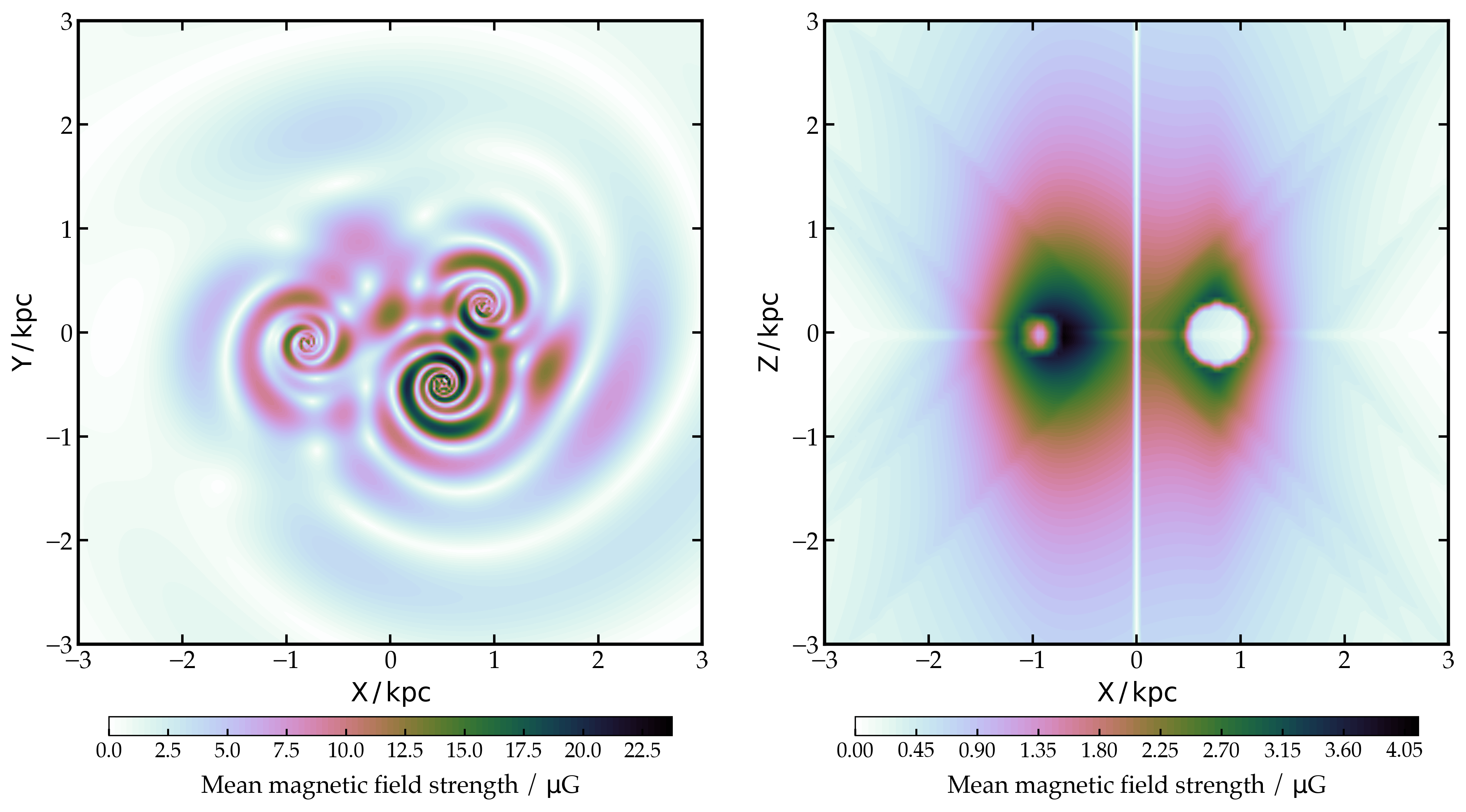}
    \caption{Mean magnetic-field strength maps used as inputs for the CRPropa simulations, shown in face-on (\textit{left panel}) and edge-on (\textit{right panel}) projections. The model consists of three logarithmic spiral components centred at APC\,1 ($0.9$\,kpc, $0.25$\,kpc), APC\,2 ($0.5$\,kpc, $-0.5$\,kpc), APC\,3 ($-0.8$\,kpc, $-0.1$\,kpc) following the star-forming regions identified by \citet{Buzzo_2021}. A spiral pitch angle of $12^\circ$ and an X-shaped halo field with an opening angle of $50^\circ$ are adopted to define the large-scale 3D field structure.}
    \label{bfield}
\end{figure*}

\subsection{Faraday rotation measure}
The rotation measure map in panel d) of Fig.~\ref{pol_meerkat}, \ref{pol_atca21}, and \ref{pol_atca55} has been corrected for the Milky Way foreground \citep{Hutschenreuter_2023}. The rotation measure value of the Galactic foreground is $3.54\pm0.92$\,rad\,m$^{-2}$, which is relatively low. Positive values indicate a magnetic field directed toward us, while negative values indicate one pointing away.
In the MeerKAT L-band data shown in panel d) of Fig.~\ref{pol_meerkat}, we observe an overall negative rotation measure as we move into the halo. Near the star-forming regions, there are rapid changes between positive and negative values, however, the relatively low resolution of our data prevents us from fully resolving these quick variations.
When analysing the ATCA L/S-band data in Fig.~\ref{pol_atca21}, we detect very rapid fluctuations between positive and negative rotation measure values. Within the star-forming cores, depolarisation limits our ability to trace the magnetic field.

However, in the regions between the star-forming knots, the rapid changes in rotation measure suggest the presence of small-scale dynamos. These small-scale dynamos are processes in which turbulent motions of the interstellar gas stretch, twist, and fold magnetic field lines, amplifying the magnetic field locally on short spatial scales \citep[e.g.][]{beck_2019, xu_2021, Pfrommer_2022}. Unlike large-scale dynamos, which produce coherent fields over kiloparsec scales, small-scale dynamos create tangled, rapidly varying magnetic structures. This could explain the observed quick variations in rotation measure between the star-forming cores, indicating strong, localised magnetic activity.
In Fig.~\ref{pol_atca55}, we do not observe any significant rotation measure, consistent with the lack of notable magnetic activity in this source.

\section{CR transport modeling}
\label{crtransport}
CRs in the merging dwarf galaxy system can be accelerated by the shocks generated during the collisions between the interstellar medium (ISM) of interacting galaxies \citep{Lisenfeld_2010, Hu_2025}, and also at the supernova remnant shocks. Transport of CRs specifically through diffusion and advection can influence the SFR and the formation of large-scale galactic wind and shape the radio intensity profile from relativistic CREs \citep{Heesen_2016, Dashyan_2020}. Therefore, the modelling of CR transport is necessary to probe the insights of magnetic field structure, advection speed and to predict the nature of diffusion processes, whether isotropic or anisotropic, based on the consistency with the observations.

\subsection{3D transport of CREs}
Since 1D modelling \citep{Heald_2022,Stein_2023} neglects the galaxy’s full 3D structure and, in edge-on systems, ignores radial features such as spiral arms, the radial structure of NGC\,1487 (inclination $i=53.4^\circ$) can still significantly affect the radio emission along the line of sight. Therefore, describing this galaxy in 3D, considering both diffusion and advection, is necessary, especially for an interacting galaxy system due to the presence of highly fluctuating density, velocity, and magnetic field \citep{Hu_2025}.
We introduce the theoretical study of 3D transport in NGC\,1487 by applying publicly available Monte-Carlo code \texttt{CRPropa}  \citep{Merten_2017, Batista_2022,Doerner_2023,Aravinthan_2025}. Although initially \texttt{CRPropa}\footnote{\url{https://crpropa.desy.de}} was developed to perform the transport of ultra-high-energy hadronic CRs by ballistic propagation through the solution of the equation of motion, \texttt{CRPropa\,3.1} is capable of executing simulations in different diffusive environments by solving the transport equation in terms of a stochastic differential equation for the diffusive propagation. The ballistic approach and diffusion module can be used within the same framework as the particle densities in a diffusive scenario can be calculated from the trajectory of pseudo-particles. To model the 3D transport of CR in NGC\,1487, we use the diffusion module where the transport equation for CREs can be represented by the diffusion-advection equation,
\begin{equation}
\frac{\partial n}{\partial t}
= D\,\nabla^2 n 
- \mathbf{u}_{\mathrm{adv}} \cdot \nabla n
- \frac{\partial}{\partial E} \left( \frac{\partial E}{\partial t} n \right)
+ S,
\label{transport_equation}
\end{equation}
where $n$ is the particle density distribution, $D$ is the diffusion coefficient, $\mathbf{u}_{\mathrm{adv}}$ is the advection speed, $\frac{\partial E}{\partial t}$ is the energy loss and $S$ is the source term, considering the isotropic diffusion. However, recent results \citep[e.g.][]{Kleimann_2025} indicate that diffusion in such systems is likely anisotropic, with diffusion coefficients differing by up to a factor of $\sim 40$ between directions parallel and perpendicular to the magnetic field. 
Here, the energy loss term represents only synchrotron emission as a continuous process. The modelling of the magnetic field in a merging system like NGC\,1487 is complex and the field structure evolves from the pre-merger to the post-merger stage as described in \citet{Rodenbeck_2016, Whittingham_2023, Hu_2025} through magneto-hydrodynamic simulations. This differs from the magnetic field of the isolated dwarf galaxy \citep{Siejkowski_2018}. 
The brightest star-forming complexes in NGC\,1487 (APC\,1, APC\,2, and APC\,3; Sect.\ref{region_sed_sect}) trace the locations where the interaction has strongly compressed the ISM. In the absence of direct constraints on the large-scale magnetic-field geometry from previous work, it is reasonable to assume multiple magnetised components originating from progenitor systems. Interacting dwarf galaxies commonly host spiral-like magnetic fields when viewed face-on and X-shaped fields when viewed edge-on \citep[][]{Heesen_2008, Heesen_2011, Bera_2019, Stein_2025}. The merger-driven overlap and distortion of such components would naturally produce the complex magnetic-field configuration shown in Fig.\ref{bfield}.

We place the center of 3 spirals at the position of APC\,1 ($0.9$\,kpc, $0.25$\,kpc), APC\,2 ($0.5$\,kpc, $-0.5$\,kpc), APC\,3 ($-0.8$\,kpc, $-0.1$\,kpc) following \citet{Buzzo_2021}. We assumed that the spiral field with a pitch angle of $12^\circ$ and the opening angle of X-shape field is $50^\circ$ as at these positions star-forming regions are located.
We structure the overlapping magnetic field strength in line with the calculated equipartition magnetic field strength at regions APC\,1, APC\,2, and APC\,3 as discussed in Sect.~\ref{equipartition_sect}.
We consider the maximum radius of NGC\,1487 is 2.75\,kpc and particles that reach this boundary exit the simulation volume. Electrons are injected at APC\,1, APC\,2, APC\,3 within the sphere of radius 0.5\,kpc, 0.1\,kpc, and 0.1\,kpc, respectively with an injection spectrum $dN / dE\propto E^{-\gamma}$ in the energy range of $0.1\,\rm GeV$ to $50\,\rm GeV$. 
At the initial stage, we use a constant diffusion coefficient $D= 10^{28}\,{\rm cm^2 s^{-1}}$, only considering the parallel diffusion and propagating electrons for 0.1\,Gyr to achieve the steady state distribution of CREs. From this simplified simulation set up, $\gamma = 2.2$ provides the overall non-thermal radio spectral index of $\alpha_{\rm nth}= -0.678$, found in NGC\,1487 (see Sect.~\ref{sed_sec}), only considering diffusion. If we consider both diffusion and advection with constant $\mathbf{u}_{\mathrm{adv}} = \mathrm{sgn}(z)\, u\, \mathbf{e}_{z}$, where $u= 160\,{\rm km\,s^{-1}}$\footnote{calculated using the rotation velocity $u=\sqrt{2} \varv_{\rm rot}$ \citep{Heesen_2018_spinnaker}}, $\alpha_{\rm nth}= -0.678$ is produced by $\gamma \sim 2.34$ which is close to equilibrium electron spectral index. This is because of the high advection speed; electrons are advected before they can diffuse in the galaxy.
This is also reflected in the calculated CRE transport timescales for diffusion and advection (see Eqs. \ref{diff} and \ref{escape} in Appendix \ref{timescale}). Using the adopted scale height of the synchrotron-emitting region, taken as half the CRE propagation path length ($h=2.75$\,kpc), diffusion coefficient $D$, and an escape velocity derived from the rotation velocity \citep[e.g.][]{Heesen_2018_spinnaker}, we obtain a diffusion timescale of $\tau_{\rm diff} = 228.17$\,Myr and an advection timescale of $\tau_{\rm adv} = 16.72$\,Myr.

Therefore, an extensive theoretical study is required to analyse the morphology of the radio emission and radio spectral index, yet this is out of the scope of this paper. Considering Kolmogorov turbulence with an anisotropic diffusion model \citep[e.g.][]{Kleimann_2025, luebke_2025}, we show that with the 3D simulation environment of CRPropa, we can use the geometry of the merging galaxies \citep{Hu_2025}, including a 3D structure of the magnetic field, to reproduce the spectral behavior as a function of the distance from the disk. This approach for the first time opens up the possibility to study 3D effects of cosmic-ray transport in complex systems like NGC\,1487. In future work (Das in preparation), we will be able to investigate the properties of cosmic-ray transport via a detailed comparison of the morphology of the galaxy with the simulations.

\section{Discussion}
\label{discussion}
\subsection{Energy losses}
\label{losses}
The spectral break observed at $6.2\pm1.3$\,GHz in NGC\,1487 in Fig.~\ref{sed} is consistent with features reported in other starburst dwarf galaxies such as NGC\,1569 and NGC\,4449 \citep{klein_2018}. This break can be understood as the frequency at which the radiative cooling time of CREs becomes comparable to their escape time from the star-forming regions. At higher frequencies, CREs enter a loss-dominated regime, where they rapidly lose energy through synchrotron radiation and/or inverse-Compton scattering before they can propagate into the halo.  

The overall non-thermal spectral index of $\alpha \approx -0.678$ ($S_\nu \propto \nu^\alpha$) provides an important diagnostic of the underlying CRE population. This index represents the transport-dominated regime below the spectral break. Using the relation  
$\alpha = (\gamma - 1)/2$,
we infer a CRE energy spectral index of $\gamma = 2.36$ ($n(E) \propto E^{-\gamma}$). This value is consistent with the initial injection spectrum, and agrees well with the typical injection index of $\gamma \approx 2.3$ inferred from supernova remnants, where diffusive shock acceleration is believed to be the main acceleration process \citep{Ruszkowski_2023}.  
In the loss-dominated regime, the steady-state CRE spectrum steepens by exactly one power, such that  
$n(E) \propto E^{-(\gamma+1)}$.
This corresponds to a steepening of the synchrotron spectral index by $\Delta \alpha = 0.5$, leading to a cooled spectral index of  $\alpha_{ \mathrm{cooled}} = \gamma/2  \approx 1.18$.
Separating the contributions of synchrotron and inverse-Compton losses is challenging, since both scale as $\dot{E}_e \propto E^2$ and thus produce similar spectral steepening. Their relative importance depends on the energy densities of the magnetic field ($u_{\rm B}$) and the ambient photon field ($u_{\rm ph}$), with inverse-Compton losses dominating when $u_{\rm ph} > u_{\rm B}$. In merger environments like NGC\,1487, the far-infrared photon field from dust-reprocessed starlight can be very strong, potentially making inverse-Compton scattering the dominant cooling process. \citet{Lacki_2010} showed that in such cases, galaxies may deviate from the linear far-infrared--radio correlation, since efficient inverse-Compton cooling reduces the energy available for synchrotron emission.  

This result highlights a key tension with the simple `calorimeter’ model. In a pure calorimeter dominated by synchrotron and inverse-Compton losses with an injection index of $\gamma \approx 2.36$, an overall spectral index of $\alpha \approx -1.18$ is expected. In contrast, our observed harder spectrum of $\alpha = -0.678$ indicates that the calorimeter model does not apply globally in NGC\,1487. The relatively hard spectrum shows that many CREs avoid significant radiative cooling inside the galaxy. This suggests that escape mechanisms, such as diffusion or streaming along magnetic field structures, play a major role in regulating the radio continuum output of this interacting system. An additional explanation arises from the galaxy’s star-formation history \citep{Buzzo_2021}, recent and bursty star-formation episodes in APC\,1, 2 and 3, inject a substantial population of high-energy electrons on timescales shorter than the effective cooling time. The resulting radio continuum is therefore a mixture of freshly accelerated electrons near star-forming regions and partially cooled electrons undergoing local escape, which together prevent the spectrum from steepening to calorimeter predictions \citep[e.g.][]{Ruszkowski_2023}.

To quantify the relative contributions of radiative losses and escape, we compute the CRE cooling times following \citet{Werhahn_2021}, \citet{taziaux_2025} and \citet{Basu_2015}, considering synchrotron, inverse-Compton, and bremsstrahlung losses. A detailed explanation of the CRE timescale equations is provided in Appendix~\ref{timescale}. We ignore the timescale for Coulomb losses, as they are primarily effective at lower CRE energies and are largely irrelevant at the higher energies associated with synchrotron-emitting CREs \citep{Petrosian_2001, Ruszkowski_2023}.
We adopt an observing frequency of 1.28\,GHz and a magnetic field strength of $12\,\upmu$G. 
 
Following Eq.~\ref{E} in Appendix~\ref{timescale}, we derive a characteristic electron energy of $E_{\rm e} = 2.58$\,GeV. The synchrotron cooling timescale of the CREs is calculated using Eq.~\ref{syn}, yielding $\tau_{\rm syn} = 22.47$\,Myr.
Inverse Compton losses are estimated by considering both the interstellar radiation field in the dwarf galaxy and the cosmic microwave background (CMB). Adopting energy densities of $U_{\rm rad} = 5.73 \times 10^{-12}$\,erg\,cm$^{-3}$ calculated from the magnetic energy density \citep{Heesen_2014} and $U_{\rm CMB} = 4.2 \times 10^{-13}$\,erg\,cm$^{-3}$ \citep{Heesen_2018_spinnaker}, we obtain an inverse Compton cooling timescale of $\tau_{\rm IC} = 28.36$\,Myr.
Bremsstrahlung losses are computed assuming a neutral gas electron density of $n_{\rm e} = 0.5$\,cm$^{-3}$, representative of the diffuse ISM in dwarf galaxies following Eq.~\ref{brems} in Appendix~\ref{timescale}. This results in a Bremsstrahlung timescale of $\tau_{\rm brem} = 79.2$\,Myr. Combining all radiative loss processes (see Eq.~\ref{cool} in Appendix~\ref{timescale}), the total cooling timescale of the CRE population is found to be $\tau_{\rm cool} = 10.59$\,Myr.
We compare this value with the CRE escape timescale, assuming advection is the dominant transport process, which is estimated to be $\tau_{\rm esc} = \tau_{\rm adv} = 16.72$\,Myr.
Furthermore, an older independent estimate of the CRE lifetime is obtained from spectral ageing. A spectral break is observed at $\nu_{\rm brk} = 6.2$\,GHz (Fig.~\ref{sed}). Using the Jaffe-Perola ageing model \citep{hughes_1991}, which assumes continuous isotropisation of electron pitch angles, we derive a spectral age of $\tau_{\rm spec} = 14.3$\,Myr following Eq.~\ref{spectral_age}. This timescale represents the time elapsed since the most recent injection of freshly accelerated CREs \citep{Heesen_2014}.

The comparison of the characteristic timescales indicates that radiative losses dominate the CRE evolution. The total cooling time ($\tau_{\rm cool} \approx 10.6$\,Myr) is shorter than both the escape time ($\tau_{\rm esc} \approx 16.7$\,Myr) and the spectral age ($\tau_{\rm spec} \approx 14.3$\,Myr), implying that CREs lose a significant fraction of their energy before escaping the system. Synchrotron and inverse Compton losses are the dominant cooling processes, reflecting the combined influence of the magnetic field and radiation field energy densities, while Bremsstrahlung losses are negligible due to the low gas density. The agreement between the cooling timescale and the spectral age supports a radiative origin of the spectral break at 6.2\,GHz and validates the use of the Jaffe-Perola ageing model.
The resulting timescales further indicate that large-scale CRE escape via galactic winds is inefficient, and that the observed radio emission is dominated by synchrotron-regulated cooling. The galaxy therefore behaves as an approximate, though not perfect, CRE calorimeter at GHz frequencies, with escape and local transport processes contributing at a secondary level.

\subsection{Role of the magnetic field}
\subsubsection{Small-scale dynamo in galaxy core}
\label{dynamo}
The rapid fluctuations in RM observed in the star-forming knots of NGC\,1487 strongly suggest that the magnetic field there is dominated by turbulent, small-scale structures. Such variability is a trend of the small-scale dynamo, which produces a tangled field amplified on the scales of turbulent driving, rather than the coherent, ordered fields expected from a large-scale dynamo \citep{beck_2019}. In this system, the small-scale dynamo is likely powered by the intense turbulence generated during the merger-driven starburst, where stellar winds and overlapping supernova explosions inject kinetic energy into the ISM. 

The efficiency of this process is demonstrated in other dwarfs. In IC\,10, the magnetic field strength correlates with the star formation rate as $B \propto \mathrm{SFR}^{0.35}$ \citep{Basu_2017}, consistent with fluctuation dynamo predictions. In NGC\,1569, the magnetic morphology is directly tied to supernova-driven H$\alpha$ bubbles, again indicating amplification through turbulence \citep{kepley_role_2010}. These observations confirm that in starbursting dwarfs, where turbulence dominates gas dynamics, the small-scale dynamo provides the primary mechanism for magnetic field growth.

The implications are twofold. Dwarf galaxies such as NGC\,1487 are close analogues of high-redshift systems, where small-scale dynamos rapidly amplify seed fields to significant strengths within $\sim10$\,Myr, ensuring early magnetisation \citep{Schleicher_2010, Rieder_2017}. At later stages, this turbulent amplification provides the seed for large-scale dynamo action, which organises the field once rotation and stratification dominate \citep{Elstner_2014}. Simulations indicate, however, that strong small-scale activity can delay large-scale ordering, extending its growth from $\sim1-2$\,Gyr to $>5$\,Gyr \citep{Schleicher_2010, Rieder_2017}.

In this context, the RM variability in NGC\,1487 is not only evidence of ongoing small-scale dynamo action, but also a snapshot of the transitional phase in galaxy assembly. Starburst-driven turbulence amplifies and sustains magnetic fields in dwarfs, while mergers and outflows transport these tangled fields into the circumgalactic medium, contributing to the magnetisation of the larger systems that these dwarfs eventually form.

\subsubsection{Magnetic field orientation in the tidal arms}
\begin{figure}
    \centering
    \includegraphics[width=\linewidth]{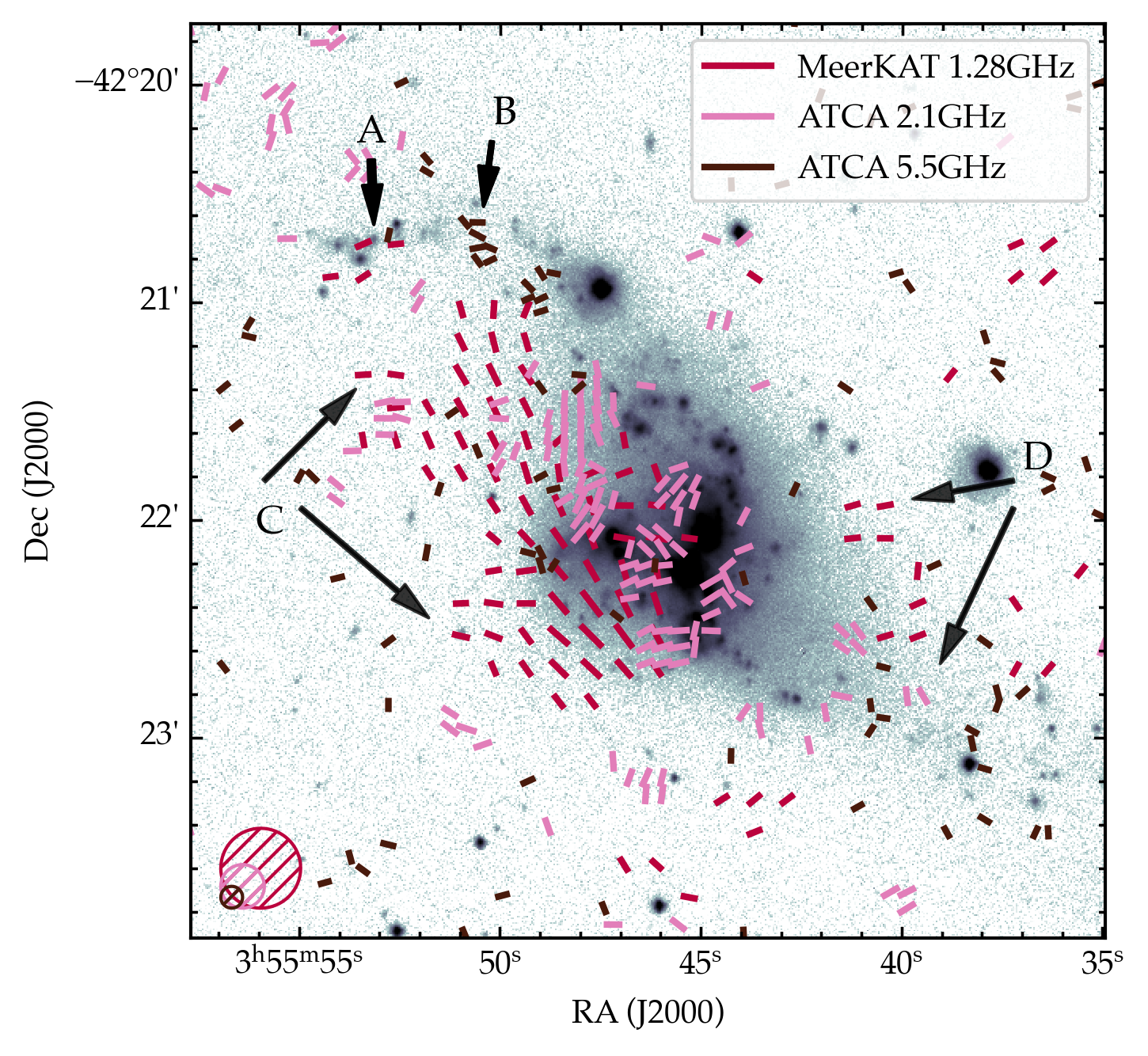}
    \caption{H$\alpha$ map with overlaid magnetic field orientation of the MeerKAT L-band data (in red), ATCA L/S-band data (in pink) and ATCA 5.5\,GHz data (in brown). The corresponding resolution is shown in left bottom corner.}
    \label{halpha_mfvec}
\end{figure}

In Fig.~\ref{halpha_mfvec}, we present the H$\alpha$ map with the overlaid magnetic field orientations derived from the different observations. The magnetic field vectors clearly extend into the tidal arm of this dwarf–dwarf merger system. We identify two main regions of interest, corresponding to the magnetic field orientations observed with MeerKAT L-band (point A in Fig.\ref{halpha_mfvec}) and ATCA C-band (point B in Fig.\ref{halpha_mfvec}). The figure shows that, after a rather chaotic field configuration in the central region of the galaxy, the magnetic field becomes more ordered and points roughly northward, toward the base of the tidal arm (as seen in the ATCA L/S-band data). Along the tidal arm, the field continues to align with the structure, reaching points B and A toward the outer end of the arm. Additionally, at points C and D in Fig.~\ref{halpha_mfvec}, the magnetic field vectors are directed toward the outer regions of the galaxy, curving slightly into the halo. This curvature suggests the presence of a large-scale magnetic field structure. These magnetic field loops are also in agreement with the simulation presented in Fig.~\ref{bfield}. At point D, the field appears to follow the tidal arm while bending outward, again indicating a possible magnetic loop extending into the halo. 

Similar results were reported by \citet{Basu_2017_interacting}, who detected ordered magnetic fields in the tidal tail of an interacting galaxy system, the Antennae galaxies. Their study showed that tidal interactions can compress and shear the ISM, which helps to align and strengthen the magnetic field on large scales. These findings, together with our results, suggest that tidal forces in galaxy mergers play an important role in shaping and maintaining large-scale magnetic field structures. Such fields can stay coherent over several kiloparsecs and are visible as polarised synchrotron emission even in the outer regions of the interacting galaxies.

\section{Conclusion}
\label{conclusion}
This study presents a comprehensive and the first radio continuum polarimetry study of the dwarf-dwarf merger NGC\,1487 to date. 
The key aspects of the study focus on investigating the magnetic field structure, CR transport, and energy loss mechanisms within this low-mass merging system using multi-band radio continuum data from MeerKAT (L-band) and ATCA (L/S-, C-, and X-bands). These can be summarised under the following points. 
\begin{enumerate}
    \item Radio continuum emission is detected all over the frequency range and is even traced in the tidal arms at MeerKAT L-band. The overall non-thermal spectral index derived from the full integrated SED is $\alpha_{\rm nth} = -0.678\pm0.085$, consistent with an aged CRE population.
    
    \item The SED analysis revealed that the best-fit model (\texttt{ffa1\_sfg\_sic}) includes free-free absorption at low frequencies, synchrotron/free-free emission, and a break due to inverse Compton losses at higher frequencies. A spectral break frequency $\nu_b \approx 6.2\pm 1.3\,\text{GHz}$ was observed. This break suggests that at frequencies above this point, CREs enter a loss-dominated regime where radiative cooling time is comparable to the escape time.

    \item Region-dependent spectral analysis shows that actively star-forming knots (e.g. APC\,2) exhibit relatively flat non-thermal spectra ($\alpha_{\rm nth} = -0.52 \pm 0.02$). In contrast, regions located at the base of the tidal arms (e.g. APC\,4) possess significantly steeper non-thermal spectra ($\alpha_{\rm nth} = -0.96 \pm 0.03$), indicating that the CREs in these outer regions have experienced substantial radiative losses or more efficient transport/escape.

    \item Calculations comparing cooling and escape timescales indicate that the CRE cooling time at 1.28\,GHz ($\tau_{\rm cool} \approx 10.6$\,Myrs) is comparable to the estimated escape time via advection in a galactic wind ($\tau_{\rm esc} \sim 16.7$\,Myrs). This suggests that radiative losses are important in regulating the CRE population, but that escape processes also play a significant role. While synchrotron losses are somewhat more efficient than inverse-Compton losses under the adopted conditions, the relatively flat observed radio spectral index ($\alpha = -0.68$) compared to the expectation for a pure electron calorimeter ($\alpha \approx -1.18$) indicates that a substantial fraction of CREs escape before losing all of their energy. The galaxy therefore does not behave as a perfect electron calorimeter; instead, its radio emission reflects a balance between radiative cooling and CRE transport.

    \item Within the star-forming knots, rapid fluctuations in RM were observed. These fluctuations strongly suggest that the core magnetic field is dominated by turbulent, small-scale structures generated by the small-scale dynamo, likely powered by intense merger-driven turbulence

    \item Outside the chaotic core, the magnetic field orientations clearly extend and align with the tidal arms, pointing roughly northward toward the base of the tidal arm. This result, consistent with findings in other interacting galaxies \citep{Basu_2017_interacting}, indicates that tidal forces play an important role in shaping and maintaining large-scale magnetic field structures in merging dwarf galaxies.

\end{enumerate}

\begin{acknowledgements}
ST, CJR, DJB, MS, JBT, SD, and RJD acknowledge the support from the DFG via the Collaborative Research Center SFB1491 \textit{Cosmic Interacting Matters - From Source to Signal} (project no.\ 445052434). PK acknowledges the support of the BMBF project 05A23PC1 for D-MeerKAT.
The MeerKAT telescope is operated by the South African Radio Astronomy Observatory, which is a facility of the National Research Foundation, an agency of the Department of Science and Innovation.
The Australia Telescope Compact Array is part of the Australia Telescope National Facility (https://ror.org/05qajvd42), which is funded by the Australian Government for operation as a National Facility managed by CSIRO. We acknowledge the Gomeroi people as the Traditional Owners of the Observatory site.
We acknowledge data storage and computational facilities by the University of Bielefeld, which are hosted by the Forschungszentrum Jülich and were funded by German Federal Ministry of Education and Research (BMBF) projects D-LOFAR IV (05A17PBA) and D-MeerKAT-II (05A20PBA), as well as technical and operational support by BMBF projects D-LOFAR 2.0 (05A20PB1) and D-LOFAR-ERIC (05A23PB1), and German Research Foundation (DFG) project PUNCH4NFDI (460248186). 
The Legacy Survey team makes use of data products from the Near-Earth Object Wide-field Infrared Survey Explorer (NEOWISE), which is a project of the Jet Propulsion Laboratory/California Institute of Technology. NEOWISE is funded by the National Aeronautics and Space Administration.
The Legacy Surveys imaging of the DESI footprint is supported by the Director, Office of Science, Office of High Energy Physics of the U.S. Department of Energy under Contract No. DE-AC02-05CH1123, by the National Energy Research Scientific Computing Center, a DOE Office of Science User Facility under the same contract; and by the U.S. National Science Foundation, Division of Astronomical Sciences under Contract No. AST-0950945 to NOAO.

\end{acknowledgements}

\bibliographystyle{aa} % style aa.bst
\bibliography{literatur} % your references Yourfile.bib

\begin{appendix}
\onecolumn
\section{Calculations for thermal correction}
\label{thermalcorrection}
Follow Eq~\ref{thcorr}, we can determine the thermal emission $S_0(\nu)$ by using the H$\alpha$ emission as its tracer, because it originates from the recombination of the same free electrons that produces the free–free emission. Therefore, we use the free--free emission $S_0$ developed by \citet{Deeg_1997} to use as thermal emission subtraction.
\begin{equation}
    \biggl(\frac{S_0(\nu)}{\rm erg \,s^{-1}\,cm^{-2}} \biggr) = 1.14 \times 10^{14}  \biggl(\frac{T_{\rm e}}{10^4\,\rm K}\biggr)^{0.034} \biggl(\frac{F_{\rm H\alpha}}{\rm erg \,s^{-1}\,cm^{-2}}\biggr)
\end{equation}
To take now into account the internal absorption in NGC\,1487, we use the free--free optical depth $\tau_{\rm ff}$ defined as 
\begin{equation}
    \tau_{\rm ff} = 0.082\,\biggl(\frac{T_{\rm e}}{1\,\rm K}\biggr)^{-1.35} \biggl(\frac{\nu}{\rm GHz}\biggr) \biggl(\frac{\rm EM}{\rm cm^{-6}\,pc}\biggr)
    \label{opticaldepth}
\end{equation}
while the emission measure EM is directly related to the H$\alpha$ flux density. 
\begin{equation}
    \begin{aligned}
        \biggl(\frac{F_{\rm H\alpha}}{\rm erg \,s^{-1}\,cm^{-2} sr^{-1}}\biggr) &= 9.41 \times  10^{-8}\,\biggl(\frac{T_{\rm e}}{10^4\,\rm K}\biggr)^{-1.017} 
      %  \\ &\quad 
        \times 10^{-0.029/\biggl(\frac{T_{\rm e}}{10^4\,\rm K}\biggr)}\biggl(\frac{\rm EM}{\rm cm^{-6}\,pc}\biggr)
    \end{aligned}
\end{equation}
With the solid angle $\Omega$, we can transform from steradian to the integrated H$\alpha$ emission with knowing that $1\, \text{sr} = 4.25 \times 10^{10}\, \text{arcsec}^2$. Therefore, we can use $\Omega = \frac{2\pi \theta^2}{4.25 \times 10^{10}\, \text{arcsec}^2}$ \citep{Smoot_1998} where $\theta$ is the angular radius in arcsec, we get 
\begin{equation}
  \biggl(\frac{F_\text{total}}{\rm erg/s/cm^2/sr^1}\biggr) = \biggl(\frac{F_\text{total}}{\rm erg/s/cm^2}\biggr) \times \biggl(\frac{\Omega}{sr}\biggr)^{-1}
\end{equation}

\section{Radio flux densities}
\label{flux_density_table_sect}
In this section, we present the values for the radio flux densities of the MeerKAT, ATCA and the literature data in Table~\ref{flux_density_table}. In Table~\ref{sed_fit_params}, we show the exact parameter values and its uncertainties of the best-fit model.

\begin{table}[]
\centering
\caption{Total flux density measurements of NGC\,1487.}
\label{flux_density_table}
\begin{tabular}{cccc}
\toprule
Frequency & $S_{\rm tot}$ & $\sigma_{\rm tot}$ & References \\
(MHz) & (mJy) & (mJy) & \\
\midrule
87.7   & 85.00 & 8.50 & GLEAM-X, \citet{gleamx} \\
118.4  & 90.00 & 9.00 & GLEAM-X, \citet{gleamx} \\
154.0  & 74.00 & 7.40 & GLEAM-X, \citet{gleamx} \\
200.0  & 74.00 & 7.40 & GLEAM-X, \citet{gleamx} \\
215.7  & 72.00 & 7.90 & GLEAM-X, \citet{gleamx} \\
%(300 & 52.00 & 10.4 & GLEAM 300 \citet{Duchesne_2025_gleam300})\\
887.0  & 35.80 & 3.58 & RACS-Low, \citet{racs} \\
943.0  & 35.70 & 3.57 & EMU, \citet{emu_main} \\
975.6  & 33.95 & 3.40 & MeerKAT, this work \\
1127.1 & 30.89 & 3.09 & MeerKAT, this work \\
1278.7 & 26.66 & 2.67 & MeerKAT, this work \\
1280.0 & 26.89 & 2.69 & MeerKAT, this work \\
1322.0 & 25.25 & 2.53 & MeerKAT, this work \\
1368.0 & 19.90 & 1.99 & ATCA, this work \\
1430.2 & 25.02 & 2.50 & MeerKAT, this work \\
1582.1 & 21.67 & 2.17 & ATCA, this work \\
1844.0 & 19.73 & 1.97 & ATCA, this work \\
2100.0 & 18.40 & 1.84 & ATCA, this work \\
2356.0 & 15.82 & 1.58 & ATCA, this work \\
2868.0 & 14.27 & 1.43 & ATCA, this work \\
4870.0 & 10.60 & 1.21 & ATCA, \citet{Grundy_2025} \\
5470.0 & 10.39 & 1.04 & ATCA, \citet{Grundy_2025} \\
5480.0 & 9.99  & 1.28 & ATCA, \citet{Grundy_2025} \\
5500.0 & 7.60  & 0.76 & ATCA, this work \\
6012.0 & 6.60  & 0.66 & ATCA, this work \\
6140.0 & 8.85  & 1.01 & ATCA, \citet{Grundy_2025} \\
8488.0 & 4.80  & 0.48 & ATCA, this work \\
9000.0 & 5.50  & 0.55 & ATCA, this work \\
9000.0 & 6.07  & 1.30 & ATCA, \citet{Grundy_2025} \\
9470.0 & 5.09  & 0.99 & ATCA, \citet{Grundy_2025} \\
9512.0 & 4.60  & 0.46 & ATCA, this work \\
9940.0 & 4.76  & 1.06 & ATCA, \citet{Grundy_2025} \\
16930.0 & 3.97 & 0.70 & ATCA, \citet{Grundy_2025} \\
\bottomrule
\end{tabular}
\end{table}

\begin{table}[]
\centering
\caption{Best-fit parameters and the fit statistics of the SED model (\texttt{ffa1\_sfg\_sic}) for NGC\,1487. Uncertainties correspond to 1$\sigma$ errors.}
\begin{tabular}{lcc}
\toprule
Parameter & Value & Uncertainty \\
\midrule
$A$ & 129.06 & 7.81 \\
$\alpha$ & $-0.678$ & 0.085 \\
$B$ & 5.36 & 1.81 \\
$\nu_1$ [MHz] & 94.82 & 20.95 \\
$\nu_{\rm b}$ [MHz] & 6220.7 & 1269.0 \\
$\Delta \alpha$ & 2.41 & 1.20 \\
$\nu_{\rm ref}$ [MHz] & 100 &  \\
\midrule
\multicolumn{3}{l}{\textbf{Fit statistics}} \\
$\chi^2$ & 28.707 &  \\
$\chi^2_{\rm red}$ & 1.104 &  \\
AIC & 8.525 &  \\
BIC & 17.320 &  \\
\bottomrule
\end{tabular}
\label{sed_fit_params}
\end{table}

\section{Uncertainty maps}
\label{uncertainties}
\begin{figure*}
    \centering
    \includegraphics[width=0.75\linewidth]{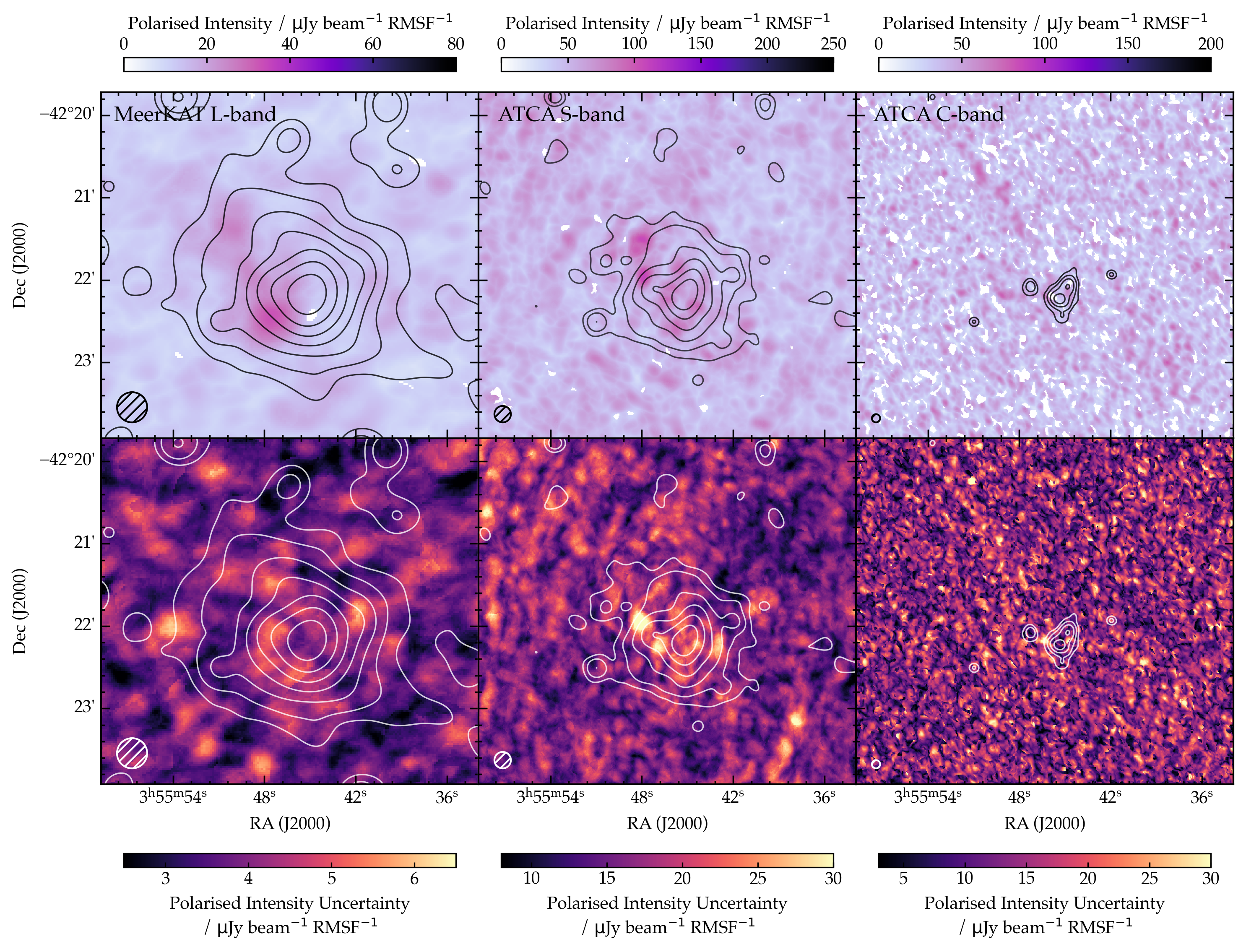}
    \caption{Polarised Intensity (top row) and its uncertainty maps (bottom row) of the MeerKAT L-band data (left panels), ATCA L/S-band (middle panels) and ATCA C-band (right panels). Total intensity of MeerKAT 1194\,MHz (left panels) ATCA 2166\,MHz (middle panels) and ATCA 5430\,MHz (right panels) contours are overlaid starting at $3\sigma$ and increase by a factor of 2 ($\sigma_{1194\,{\rm MHz}}= 30\upmu$Jy/beam; $\sigma_{2166\,{\rm MHz}}= 15\upmu$Jy/beam; $\sigma_{5430\,{\rm MHz}}= 11\upmu$Jy/beam). For each panel, the circular beam appears in the lower left panel.}
    \label{pi_uncert}
\end{figure*}

\begin{figure*}
    \centering
    \includegraphics[width=0.75\linewidth]{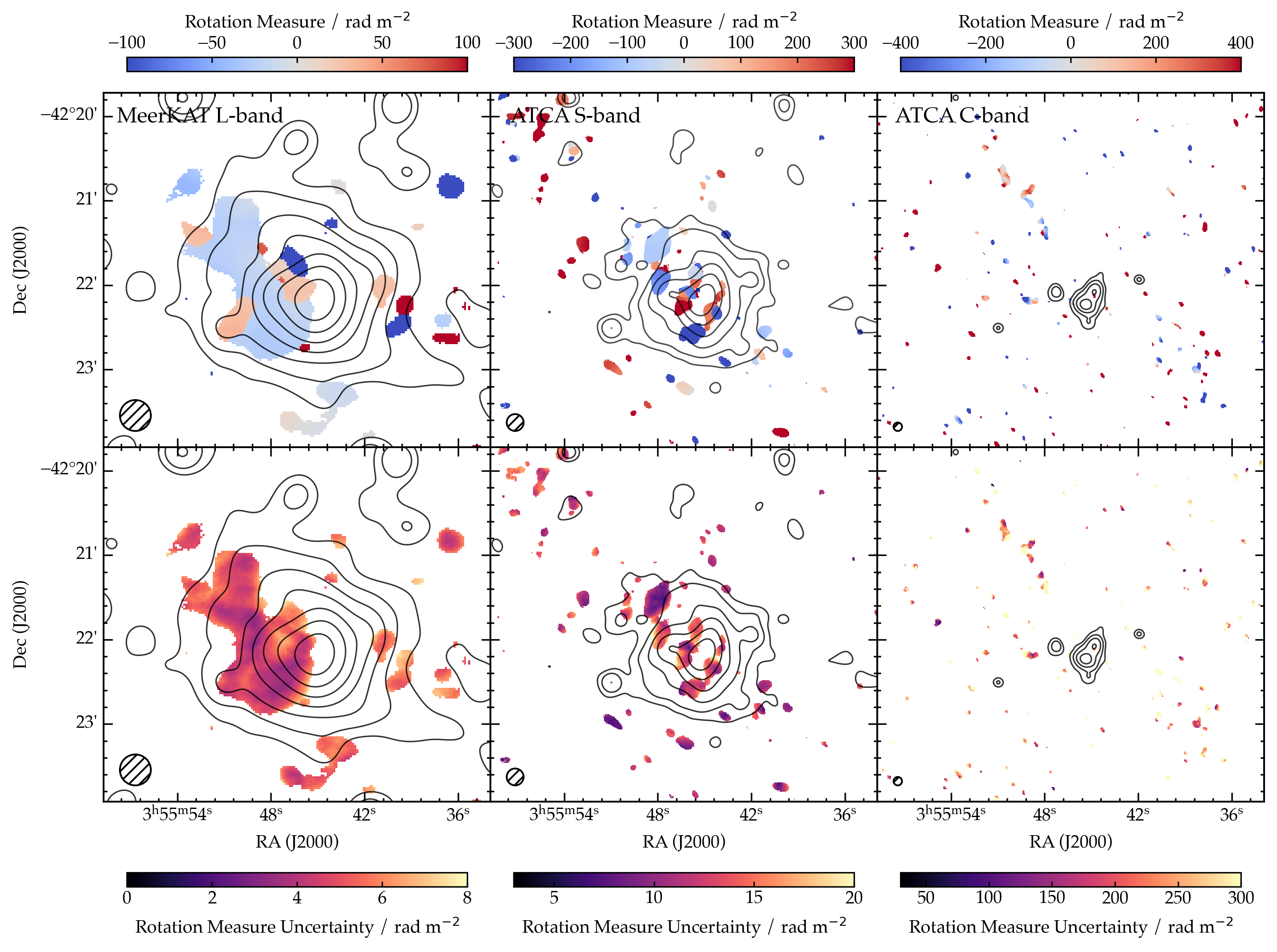}
    \caption{Foreground-corrected rotation measure (top row) and its uncertainty maps (bottom row) of the MeerKAT L-band data (left panels), ATCA L/S-band (middle panels) and ATCA C-band (right panels). Total intensity of MeerKAT 1194\,MHz (left panels) ATCA 2166\,MHz (middle panels) and ATCA 5430\,MHz (right panels) contours are overlaid starting at $3\sigma$ and increase by a factor of 2 ($\sigma_{1194\,{\rm MHz}}= 30\upmu$Jy/beam; $\sigma_{2166\,{\rm MHz}}= 15\upmu$Jy/beam; $\sigma_{5430\,{\rm MHz}}= 11\upmu$Jy/beam). For each panel, the circular beam appears in the lower left panel.}
    \label{rm_uncert}
\end{figure*}

In this section, we show the uncertainty maps for the polarised intensity (Fig.\ref{pi_uncert}), and the rotation measure (Fig.~\ref{rm_uncert}).

\section{CRE timescales}
\label{timescale}
This section helps understanding the calculated CRE timescales in Sect.~\ref{losses}. The following equation are based on \citet{Werhahn_2021} and \citet{Basu_2015}.
For the electron energy, we follow \citet{Basu_2015}, and it can be written as 
\begin{equation}
\label{E}
\frac{E}{\rm GeV} =  \biggl[ 7.9\times \biggl(\frac{\nu_{\rm obs}}{\rm GHz}\biggr)^{1/2} \biggl(\frac{B_\perp}{\upmu {\rm G}}\biggr)^{-1/2} \biggr]     
\end{equation}
with assuming $B_\perp=B$, being the magnetic field strength in the plane of the sky, and $\nu_{\rm obs}$ being the MeerKAT L-band observing frequency.
Using the electron energy, we calculate the synchrotron timescale $\tau_{\rm syn}$ for CREs emitting at a radio frequency $\nu_{\rm obs}$
\begin{equation}
    \frac{\tau_{\rm syn}}{\rm Myrs}= 8.35\times 10^3 \biggl(\frac{E}{\rm GeV}\biggr)^{-1} \,\biggl(\frac{B}{\upmu {\rm G}}\biggr)^{-2}
    \label{syn}
\end{equation}
The inverse Compton timescale can be calculated using the radiation field and cosmic microwave background energy density. The cosmic microwave background energy density is $U_{\rm CMB} = 4.2\times 10^{-13}$\,erg\,cm$^{-3}$ \citep{Heesen_2018_spinnaker} and we assume that the radiation energy density is dominated by the magnetic energy density $U_{\rm rad} = B^2/8\pi$ \citep{Heesen_2014}. The inverse Compton timescale can be given by 
\begin{equation}
\label{ic}
        \frac{\tau_{\rm IC}}{\rm Myrs}= 4.5\times 10^2 \biggl(\frac{E}{\rm GeV}\biggr)^{-1} \,\biggl(\frac{U_{\rm rad}+ U_{\rm CMB}}{10^{-12} {\rm erg\,cm^{-3}}}\biggr)^{-1}
\end{equation}
Bremsstrahlung timescale depends on the average number density of total neutral gas $\langle n \rangle$  in the ISM, and is given by 
\begin{equation}
\label{brems}
        \frac{\tau_{\rm brems}}{\rm Myrs}= 39.6 \,  \biggl(\frac{\langle n \rangle }{\rm cm^{-3}}\biggr)^{-1} 
\end{equation}
Coulomb interaction losses become typically more significant and effective at lower CRE energies \citep{Petrosian_2001, Ruszkowski_2023}, therefore ignored in our analysis. The cooling timescale for a continous injection of CREs can be calculated by combining all the timescales
\begin{equation}
\label{cool}
        \tau_\text{cool}^{-1} =  \tau_\text{syn}^{-1} + \tau_\text{IC}^{-1} + \tau_\text{brems}^{-1}
\end{equation}

For the escape timescale $\tau_{\rm esc},$ we distinguish between two CRE transport processes, namely diffusion and advection. For each process, a characteristic timescale can be calculated to identify the dominant transport mechanism, following \citep{Werhahn_2021}. The diffusion timescale of CREs is given by

\begin{equation}
\label{diff}
\tau_{\rm diff} = \frac{h^{2}}{D},
\end{equation}

where $h$ is the CRE scale height and $D$ is the diffusion coefficient. The timescale for CREs escaping the system via advection is expressed as

\begin{equation}
\label{escape}
\tau_{\rm adv} \sim \frac{h}{\varv_{\rm esc}},
\end{equation}

where $\varv_{\rm esc}$ is the escape velocity. The escape velocity is estimated from the rotation velocity as $\varv_{\rm esc} = \sqrt{2}\, \varv_{\rm rot}$ \citep[e.g.][]{Heesen_2018_spinnaker}.

Furthermore, a spectral analysis is done following \citet{Heesen_2014}, to estimate the spectral age based on the magnetic field and the frequency of the break $\nu_{\rm brk}$ in the spectrum to calculate the time hat has elapsed since freshly accelerated CRe were injected. The spectral age are given by 
\begin{equation}
\label{spectral_age}
            \frac{\tau_{\rm spec}}{\rm Myrs}= \sqrt{\frac{2.53\times 10^{3} \biggl(\frac{B}{10\upmu {\rm G}}\biggr)}{\nu_{\rm brk} \, \biggl[\biggl(\frac{B}{10\upmu {\rm G}}\biggr)^2+ \biggl(\frac{B_{\rm CMB}}{10\upmu {\rm G}}\biggr)^2\biggr]^2}}
\end{equation}
with the CMB magnetic field strength being defined as  $B_{\rm CMB}/\upmu {\rm G} = 3.2(1 + z)^2$.

\end{appendix}

%\bsp	% typesetting comment
\label{lastpage}
\end{document}